\documentclass[sigconf]{acmart}

\usepackage{multirow}
\usepackage{algorithm,algcompatible,amsmath}
\usepackage{algpseudocode}
\usepackage{algorithmicx}
\usepackage{booktabs}
\usepackage{fancyhdr}

\copyrightyear{2022}
\acmYear{2022}
\setcopyright{rightsretained}
\acmConference[KDD '22]{Proceedings of the 28th ACM SIGKDD Conference on Knowledge Discovery and Data Mining}{August 14--18, 2022}{Washington, DC, USA}
\acmBooktitle{Proceedings of the 28th ACM SIGKDD Conference on Knowledge Discovery and Data Mining (KDD '22), August 14--18, 2022, Washington, DC, USA}
\acmDOI{10.1145/3534678.3539172}
\acmISBN{978-1-4503-9385-0/22/08}

\begin{document}

\title{Multiwave COVID-19 Prediction from Social Awareness\\using Web Search and Mobility Data}

\author{Jiawei Xue}
\email{xue120@purdue.edu}
\orcid{0000-0001-7519-6130}
\affiliation{
  \institution{Purdue University}
  \streetaddress{550 Stadium Mall Drive}
  \city{West Lafayette}
  \state{IN}
  \country{USA}
  \postcode{47907-2051}
}

\author{Takahiro Yabe}
\email{tyabe@mit.edu}
\orcid{0000-0001-8967-1967}
\affiliation{
  \institution{Massachusetts Institute of Technology}
  \streetaddress{550 Stadium Mall Drive}
  \city{Cambridge}
  \state{MA}
  \country{USA}
}

\author{Kota Tsubouchi}
\email{ktsubouc@yahoo-corp.jp}
\affiliation{
  \institution{Yahoo Japan Corporation}
  \city{Tokyo}
  \country{Japan}
}

\author{Jianzhu Ma*}
\email{majianzhu@pku.edu.cn}
\orcid{0000-0002-8236-6609}
\affiliation{
 \institution{Peking University; Beijing Institute for General Artificial Intelligence}
 \city{Beijing}
 \country{China}}

\author{Satish V. Ukkusuri*}
\orcid{0000-0001-8754-9925}
\email{sukkusur@purdue.edu}
\affiliation{
  \institution{Purdue University}
  \city{West Lafayette}
  \state{IN}
  \country{USA}}

\begin{abstract}
Recurring outbreaks of COVID-19 have posed enduring effects on global society, which calls for a predictor of pandemic waves using various data with early availability. Existing prediction models that forecast the first outbreak wave using mobility data may not be applicable to the multiwave prediction, because the evidence in the USA and Japan has shown that mobility patterns across different waves exhibit varying relationships with fluctuations in infection cases. Therefore, to predict the multiwave pandemic, we propose a Social Awareness-Based Graph Neural Network (SAB-GNN) that considers the decay of symptom-related web search frequency to capture the changes in public awareness across multiple waves. Our model combines GNN and LSTM to model the complex relationships among urban districts, inter-district mobility patterns, web search history, and future COVID-19 infections. We train our model to predict future pandemic outbreaks in the Tokyo area using its mobility and web search data from April 2020 to May 2021 across four pandemic waves collected by Yahoo Japan Corporation under strict privacy protection rules. Results demonstrate our model outperforms state-of-the-art baselines such as ST-GNN, MPNN, and GraphLSTM. Though our model is not computationally expensive (only 3 layers and 10 hidden neurons), the proposed model enables public agencies to anticipate and prepare for future pandemic outbreaks. 
\end{abstract}



\maketitle

\section{Introduction}
\begin{figure}[t]
    \centering
    \includegraphics[width=0.90\linewidth]{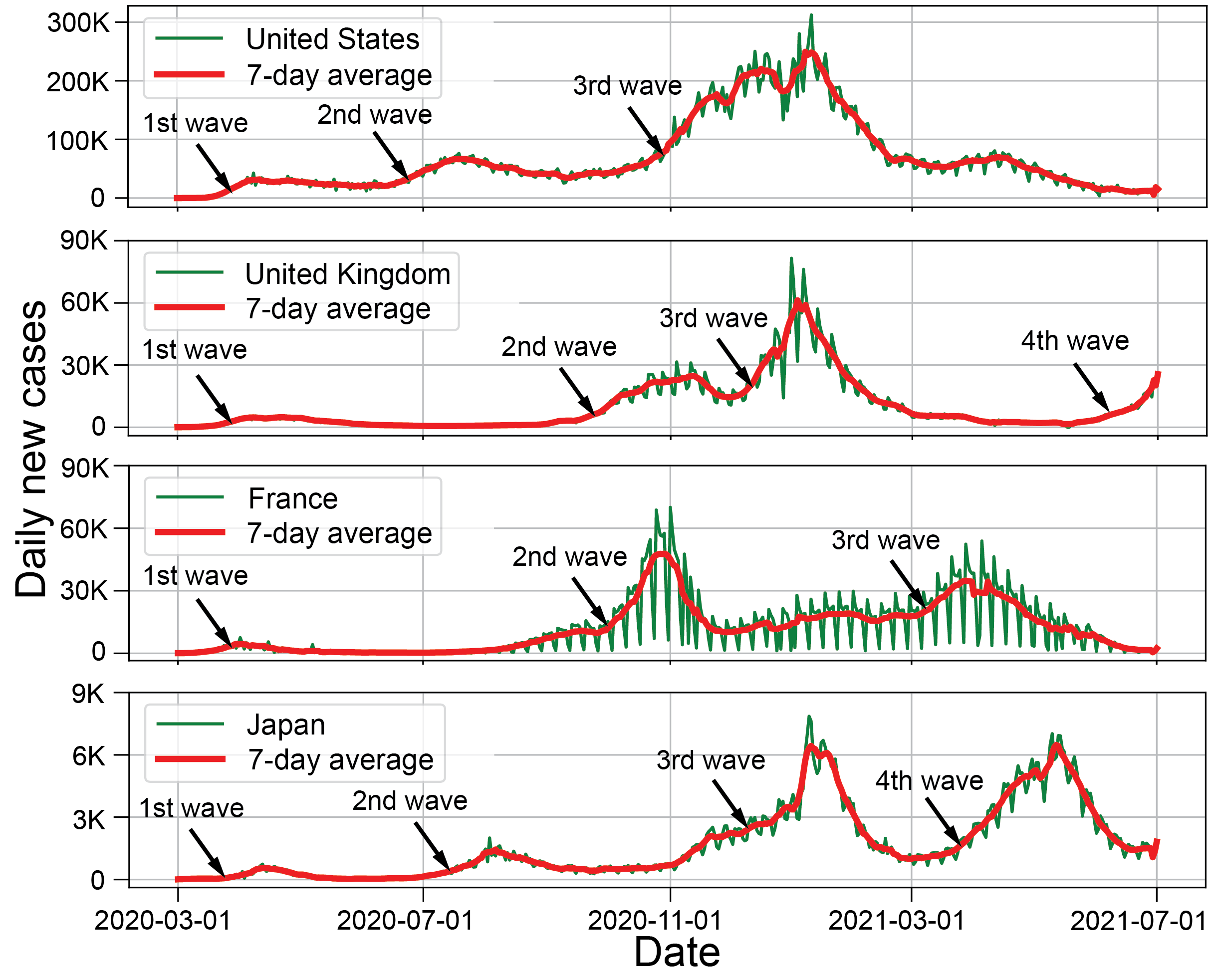}
    \vspace{-0.3cm}
    \caption{The daily new infection cases of COVID-19 for four countries \cite{covid_google}. We observe multiple pandemic waves with varying starting days and magnitudes.}
    \label{four_country}
    \vspace{-0.4cm}
\end{figure}

The spreading mechanism of COVID-19 is complicated due to its dependency on disease features and social factors such as human mobility \cite{yabe2020non,qian2021scaling}, public awareness \cite{qian2020modeling}, and intervention policies. One prominent phenomenon of the complex disease spreading process is the multiple outbreak wave, which implies the periodic rebound to a large number of infection cases \cite{leung2020first} and is obvious in many countries such as the USA, UK, France, and Japan (Figure~\ref{four_country}). Abrupt and uncertain disease outbreaks disturb individuals' daily life, government's reopening policies \cite{chang2020mobility}, medical resources managements \cite{moghadas2020projecting}, and risk assessment \cite{ye2020alpha}. It has an enormous social impact to investigate and construct an accurate model to predict the multiple waves by fully utilizing different types of data \cite{xiao2021c}. 

Many prediction models for the first outbreak wave have been proposed to anticipate the infection and death cases \cite{shahid2020predictions,chimmula2020time,kufel2020arima,schwabe2021predicting}. One critical input for these models is the mobility data \cite{kang2020multiscale,huang2020twitter,chang2021supporting}, which describes population movements and is positively related to the disease infections \cite{xiong2020mobile}. Nevertheless, continuous tracking of human mobility dynamics shows that the mobility strength did not exhibit consistent relationships with infection cases: (1) the USA: human mobility fluctuated around 95\% of the normal period from July 1 to Dec. 1, 2020 \cite{usa_covid} which witnessed the second wave in July and the third wave in November (Figure~\ref{four_country}); (2) Tokyo: the social contact index resumed to the normal level and decreased slightly after July 2020, but Tokyo experienced the second wave since then \cite{yabe2021early}. The research community has also noticed and discussed the limitation of mobility data in long-term multiwave infection prediction \cite{levin2021insights, badr2021limitations}. The inconsistency in the mobility and infection necessitates other data that is more representative of disease waves.

During COVID-19, many text-based methods have been proposed to aid communities such as to understand human's emotion states \cite{ju2021dr} and to answer peoples' questions \cite{yan2021multilingual}. Web search records collected by the web service provider have extensive applications such as customer behavior analysis \cite{goel2010predicting}, disease outbreak monitoring \cite{hisada2020surveillance}, and evacuation prediction \cite{yabe2019predicting}. For COVID-19, symptom-related web search records (e.g., \textit{fever}, \textit{cough}, and \textit{headache}) reflect the public's virus-induced symptoms that cannot be mined from the mobility data. Evidence in Tokyo showed that the Pearson correlation between High-Risk Users (which are defined from web search records) and infection cases was 0.719 with a 16-day lag for the second wave \cite{yabe2021early}. It also found people's symptom-related web search frequency was smaller during the second wave than the first wave, though the number of patients were significantly higher. These results inspire us to leverage the web search data and recover human awareness decay to predict the multiwave pandemic.

In this study, we propose a multiwave infection prediction approach\footnote{The code and data of this study is open to public and can be found at https://github.com/JiaweiXue/MultiwaveCovidPrediction.}, with the direct application as urban district-level disease outbreak early warning. District-level disease prediction has the following three requirements: (1) comprehensive data sources, such as people movement and social responses, should be included to contain various hints that are closely related to the disease spreading; (2) spatial and temporal disease transmission patterns of COVID-19 should be taken into consideration; (3) it captures complex dependency between infection cases and other factors. To deal with these challenges, we first define the \textbf{Web search Mobility Network (WMN)} whose nodes and edges maintain web search frequency and inter-district human flow information, respectively. Afterward, we propose a \textbf{Social Awareness-Based Graph Neural Network (SAB-GNN)} architecture upon the WMN to capture the spatio-temporal infection case dynamics in different urban districts. We train and test the model using real-world infection, human mobility, and web search data in Tokyo from April 2020 to May 2021, and obtain better prediction performance than state-of-the-art baseline models. Our method has three contributions: 
\begin{itemize}
\item We focus on predicting multiwave disease outbreaks that are globally prevalent but are seldom investigated. Different from a single wave prediction, multiwave prediction enables the public agencies to evaluate the long-term risk and take appropriate actions at different pandemic stages.
\item We propose the SAB-GNN by fusing historical infection, mobility, and web search data that provide sufficient evidence of potential disease outbreaks. The spatial module, temporal module, and social awareness module take separate responsibilities and jointly contribute to the prediction.
\item The proposed method is implemented on a mega-city, Tokyo, with a period spanning more than one year across four pandemic waves. We conduct a comprehensive analysis of disease outbreaks and prediction results from different models at different time intervals, which promotes a more nuanced understanding of the disease waves. 
\end{itemize}

\section{Related Work}
\subsection{Time Series Models} 
Existing COVID-19 time series models cover multiple types such as auto-regressive integrated moving average (ARIMA) \cite{kufel2020arima}, and long short-term memory (LSTM) \cite{chimmula2020time,jo2020condlstm}. Moreover, biologists and engineering scientists focus on the relationship between fatality rate and biochemical indicators \cite{zhou2020not}, human mobility \cite{jo2020condlstm}. When it comes to urban district-level disease infection prediction, while the inter-district connections provide crucial pathways for both human movement and disease dissemination, these models are insufficient to capture such spatial dependency between different urban districts. In fact, the spatial dependency information enables us to deal with the data scarcity issue which may occur during the pandemic season. For instance, assume that we have sufficient mobility and social media data for district $\mathcal{A}$ and deficient data for district $\mathcal{B}$, and recognize strong mobility connections between districts $\mathcal{A}$ and $\mathcal{B}$. Considering the connections between the two districts helps to fully utilize the infection information and predict the infection cases for both districts $\mathcal{A}$ and $\mathcal{B}$.

\subsection{Graph Neural Networks} Graph neural network (GNN) is an innovative neural network that captures the relationship between multi-hop neighborhood nodes via the message passing mechanism \cite{zhou2020graph}. In the past years, various GNN models such as graph convolutional network (GCN) \cite{kipf2016semi}, GraphSage \cite{hamilton2017inductive}, graph attention network (GAT) \cite{velivckovic2017graph} were developed and applied to many fields such as neural machine translation \cite{bastings2017graph}, visual question answering \cite{narasimhan2018out}, traffic prediction \cite{yao2019revisiting,zhang2021traffic,chen2019gated,peng2020spatial}, and network metric generation \cite{xue2022quantifying}. 

Researchers have harnessed the superiority of the GNN in modeling spatial dependency to perform the disease infection case prediction. Most published GNN approaches \cite{kapoor2020examining,panagopoulos2021transfer,gao2021stan} focused on the infection prediction before July 2020 when the first global outbreak occurred using the historical infection and mobility data. The mobility data was sufficient to reflect the disease spreading patterns during the first wave thanks to its simple relationship with the infection. First, areas attracting more passengers had higher risks of experiencing rapidly increasing cases than some lonely areas at the beginning of the first wave. Second, the travel restriction policies during the first wave suppressed the mobility strength and thus decelerated the disease outbreak \cite{xiong2020mobile}. 

Nevertheless, the interaction between infection and mobility evolved into a much more complicated status during the later waves because the infection cases were affected by a large variety of causes such as mask policy, the vaccination, which hindered the ability of raw mobility data to reflect the infection tendency. Besides, many studies have recognized that the available mobility data for COVID-19 infection cases prediction was limited by the period length \cite{panagopoulos2021transfer,rodriguez2021steering}, which resulted in unstable learned models. In summary, the long-term multiwave infection prediction requests alternative data sources that provide sufficient interconnections with the infection case under the dynamic environment. In this study, we turn to the novel web search data, which directly reveals human's awareness to the disease and potential symptoms \cite{yom2022providing}, to perform the multiwave disease prediction.
\vspace{-0.0cm}

\begin{figure*}[ht]
    \centering
    \includegraphics[width=0.72\linewidth]{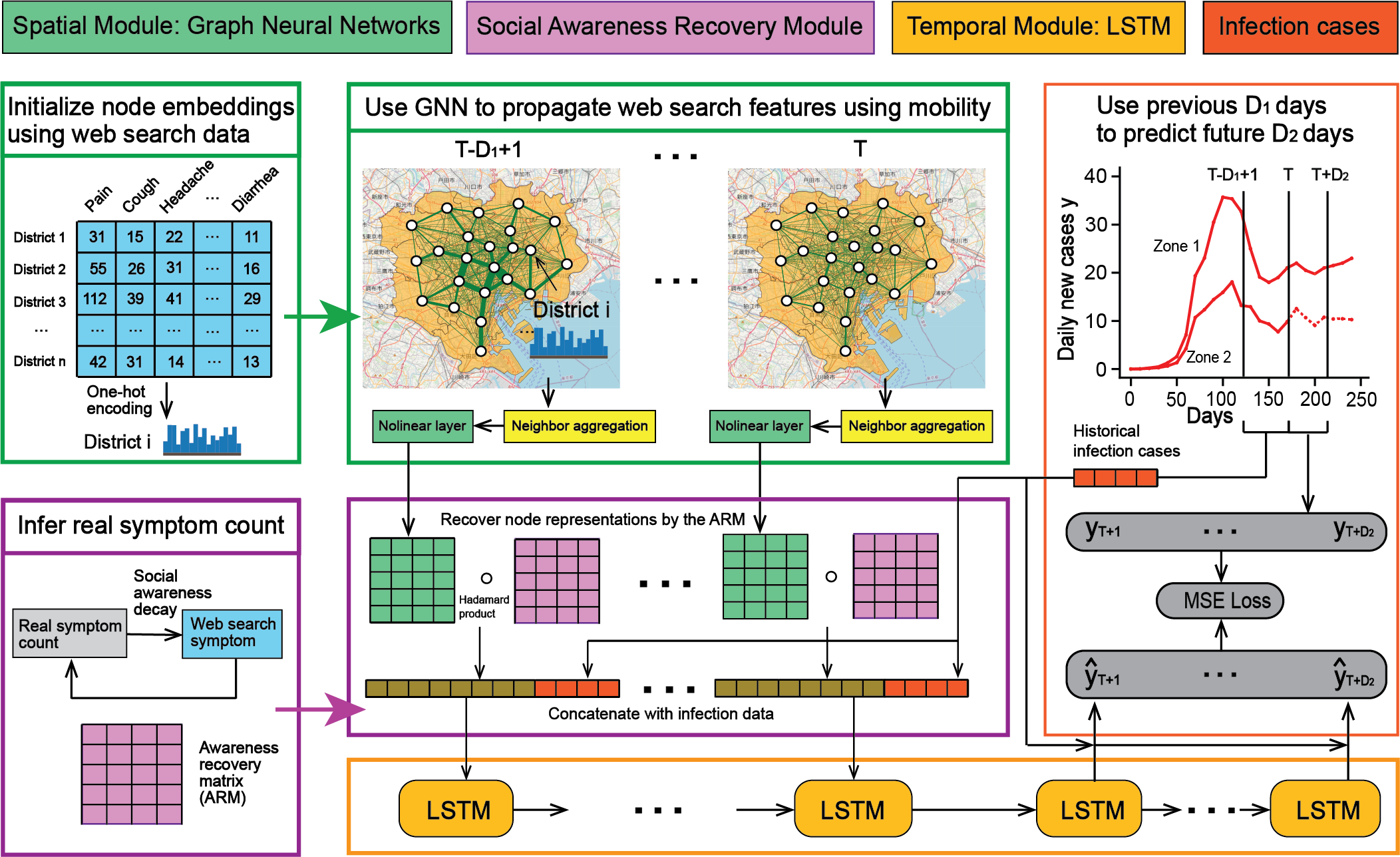}
    \caption{The framework of SAB-GNN. We first model urban districts as nodes, and propagate the web search frequency embedding using the mobility information for the last $D_{1}$ days (i.e., Spatial Module). Next, we use a learnable social awareness matrix to recover node representations (i.e., Social Awareness Recovery Module), and feed them into an LSTM sequence to predict the next $D_{2}$ days' infection cases (i.e., Temporal Module).}
    \label{figure2}
    \vspace{-0.1cm}
\end{figure*}

\section{Preliminaries}
We first describe the mobility and web search data used as features in the prediction, and then formally define our prediction task. Assume an urban area is divided into $n$ pre-defined urban districts.

\textbf{Mobility feature:} for mobile phone users within the urban area, we collect location records (longitude, latitude, and time) with the temporal resolution as around 30 minutes and spatial error as at most 100 meters. We extract each individual's trajectory points as a sequence. Next, we project the sequence points of each mobile phone user into the urban districts to obtain the number of daily inter-district trips. Note that inter-district trips build natural pathways to transmit disease viruses in the city so that these districts have correlated exposure to disease outbreak risks.

\textbf{Web search feature:} we scope $n_{w}$ COVID-19 symptom-related web search words from multiple official sources \cite{cdc,sarker2020self,wang2021covid}. We assign each mobile phone user to the urban district where his/her home is located and then aggregate the daily web searching frequency for all users living in this district. In summary, we obtain the number of inter-district trips between every two districts, and symptom-related web search frequency in each district. 

\textbf{Network definition:} We define a sequence of WMNs: $G_{t}=(V,\mathbf{E}_{t},\mathbf{H}_{t},\mathbf{I}_{t})$. Here, the subscript $t$ implies a day. $V=\{v_{1},v_{2},...,v_{n}\}$ represents $n$ urban districts. $\mathbf{E}_{t}\in \mathbb{R}^{n \times n}$ is a matrix where the ($i,j$)-entry represents the number of trips from the district $v_{i}$ to $v_{j}$ on the day $t$. We denote the daily web search frequency of residents living in the district $v_{i}$ as a vector $\mathbf{h}_{t}^{v_{i}} \in \mathbb{R}^{n_{w}}, 1\leq i\leq n$, so that $\mathbf{H}_{t} = [\mathbf{h}_{t}^{v_{1}},\mathbf{h}_{t}^{v_{2}},...,\mathbf{h}_{t}^{v_{n}}]^{T}\in \mathbb{R}^{n \times n_{w}}$ is the node feature matrix encoding the web search frequency for all districts. $\mathbf{I}_{t}\in \mathbb{R}^{n \times 1}$ is the collection of daily new infection cases in the $n$ different districts. 

\textbf{Prediction task:} Our goal is to utilize the infection, mobility, and web search data for the past $D_{1}$ days to predict new infection cases for the next $D_{2}$ days at the urban district level. On the day $T$, given $G_{t}$ for all $t \in [T-D_{1}+1, T]$, predict $\mathbf{I}_{t}$ for all $t \in [T+1, T+D_{2}]$.

\section{Social Awareness-Based Graph Neural Networks}
In this section, we first present the intuition of the SAB-GNN model, and then sequentially introduce its three modules: the spatial information propagation module, the social awareness recovery module, and the temporal information passing module, and finally declare the loss function.

We deduce that the future reported district-level infection cases are jointly influenced by the existing infection cases, the number of susceptible individuals that may have been infected (which can be mined from public's symptom-related web search data), and in-person contact patterns across the city (which is reflected by the inter-district trip numbers) and use them as features. Given that the pandemic spreading is indeed a complicated temporal process embedding on the space, we propose to build temporal and spatial modules \cite{yao2018deep} to track the infection dynamics. Lastly, the existing study finds that the public's symptom-related web search frequency decreases from the first wave to the second wave, which reveals a prevalent social phenomenon that people's awareness of a hot topic gradually declines (which is referred to as \textit{social awareness decay} in this study) \cite{yabe2021early}. Based on the fact that people adapt themselves to the mask policies, travel restrictions, and routine testings and pay less attention to the COVID-19, we propose a social awareness recovery module in the SAB-GNN to estimate the actual occurrence of COVID-19-related symptoms. In summary, we build an integrated future infection case prediction model with three modules (i.e., spatial module, awareness recovery module, and temporal module) by fusing the historical infection, mobility, and web search data. 

\subsection{Spatial Module: Graph Neural Networks}
Recall that the web search frequency vector $\mathbf{h}_{t}^{v_{i}}$ reflects the number of potential infected individuals in the urban district $v_{i}$ and $\mathbf{E}_{t}$ records the daily inter-district trips. We therefore perform the convolution operation on $\mathbf{h}_{t}^{v_{i}}$ using the $\mathbf{E}_{t}$ information under the graph neural network framework to capture the disease risk propagation properties. As shown in Figure~\ref{figure2}, using the symptom-related web search frequency in each urban district, we employ the one-hot encoding to initialize the representation for each urban district (i.e., each node in $G_{t}$) as the input matrix:  $\mathbf{X}_{t}^{(0)} = \mathbf{H}_{t}$. Following the GCN model \cite{kipf2016semi}, we define the node representation propagation rule between the layers $k$ and $(k+1)$ as:
\begin{equation}
    \mathbf{X}_{t}^{(k+1)} = \sigma(\tilde{\mathbf{D}_{t}}^{-\frac{1}{2}}\tilde{\mathbf{E}_{t}}\tilde{\mathbf{D}_{t}}^{-\frac{1}{2}}\mathbf{X}_{t}^{(k)}\mathbf{W}^{(k)}),
    \label{equ1}
\end{equation}
where 
\begin{equation}
    \tilde{\mathbf{E}_{t}} = \mathbf{E}_{t} + \mathbf{I}_{n \times n}, \tilde{\mathbf{D}_{ii}} = \sum_{j=1}^{n}\tilde{\mathbf{E}_{ij}},
    \label{equ2}
\end{equation}
and $\mathbf{W}^{(k)}$ is a learnable weight matrix, $\mathbf{I}_{n \times n}$ is the $n$ by $n$ identity matrix, $\sigma(\cdot)$ is the activation function ReLU. 

Note that we normalize the matrix $\mathbf{E}_{t}$ such that the sum of each column is equal to 1 (i.e., the sum of incoming edges on one node is 1), which is used in the existing study \cite{panagopoulos2021transfer}. In practice, it is possible to replace the spectral convolution with other GNN variants such as GAT \cite{velivckovic2017graph}, GraphSage \cite{hamilton2017inductive}. We also implement them and find quite approximate prediction performance as the GCN. The outcome of the spatial module is a matrix
\begin{equation}
    \mathbf{H}_{t}^{S}=\mathbf{X}_{t}^{(K)}=[\mathbf{x}_{t}^{v_{1}},\mathbf{x}_{t}^{v_{2}},...,\mathbf{x}_{t}^{v_{n}}]^{T},
\end{equation}
with $n$ rows that encode the web search frequency and mobility where $K$ is the number of layers.

\subsection{Social Awareness Recovery Module}
The symptom-related web search frequency is positively related to the number of actual symptom occurrences among the population \cite{yabe2021early}. As mentioned earlier, the social awareness decay effect informs that probability of symptom-related word searching gradually decreases with time after the first COVID-19 wave. To estimate the actual symptom occurrences, we propose to multiply the observed web search representation by a monotonically increasing function with respect to the time. 

Specifically, we first linearly normalize each entry of the web search record vector $\mathbf{h}_{t}^{v_i}$ to 0 and 1 by the maximal and minimal web search frequency of each word across all days in the urban district $v_{i}$, and feed them into the spatial module, and obtain $\mathbf{H}_{t}^{S}$. Next, we define an increasing function $r(t|i,t_{0})=e^{\lambda_{i}^{2}(t-t_{0})}$ regarding $t$ to recover the social awareness:
\begin{equation}
   \tilde{\mathbf{x}}_{t}^{v_{i}} = \mathbf{x}_{t}^{v_{i}}r(t|i,t_{0}) =  \mathbf{x}_{t}^{v_{i}}e^{\lambda_{i}^{2}(t-t_{0})},
   \label{equ3}
\end{equation}
where $\lambda_{i}^{2}$ is learnable and measures the social awareness recovery rate in $v_{i}$. $t$, $t_{0}$ represent the current day and the first day of the study period, respectively. Given that the land use, economy type, demographic characteristic discrepancy may lead to spatially varying social awareness rates, we specify district-dependent $\lambda_{i}^{2}$ to encode its unique awareness decay behavior. A large value of $\lambda_{i}^{2}$ implies that social awareness of COVID-19 for residents living in $v_{i}$ declines rapidly and we therefore adopt this large value to recover the social awareness. Note that we introduce the square in $\lambda_{i}^{2}$ to ensure that it is non-negative. Collectively speaking, the social awareness recovery module transforms $\mathbf{H}_{t}^{S}$ to $\tilde{\mathbf{H}}_{t}^{S}$ by
\begin{equation}
    \tilde{\mathbf{H}}_{t}^{S} = \mathbf{H}_{t}^{S}\circ \mathbf{M}_{t,t_{0}},
    \label{equ4}
\end{equation}
where $\mathbf{M}_{t,t_{0}}$ is the awareness recovery matrix (ARM):
\begin{equation}
    \mathbf{M}_{t,t_{0}} =
    \begin{bmatrix}
        e^{\lambda_{1}^{2}(t-t_{0})} & e^{\lambda_{1}^{2}(t-t_{0})}& ... & e^{\lambda_{1}^{2}(t-t_{0})} \\
        e^{\lambda_{2}^{2}(t-t_{0})} & e^{\lambda_{2}^{2}(t-t_{0})}& ... & e^{\lambda_{2}^{2}(t-t_{0})} \\
        \vdots & \vdots  & \ddots & \vdots\\ 
        e^{\lambda_{n}^{2}(t-t_{0})} & e^{\lambda_{n}^{2}(t-t_{0})}& ... & e^{\lambda_{n}^{2}(t-t_{0})} \\
    \end{bmatrix},
    \label{equ5}
\end{equation}
and $\circ$ is the Hadamard product. This transform implies that the entries in $\mathbf{H}_{t}^{S}$ are amplified to capture the actual disease risk which is underestimated due to the social awareness decay effect when $t$ is large. In the end, we perform 0-1 normalization on the infection matrix $\mathbf{I}_{t}$ to obtain $\tilde{\mathbf{I}}_{t}$, and arrive at the output of the social awareness recovery module by concatenating the web search and infection representations, which is
\begin{equation}
    \mathbf{H}_{t}^{A}=[\tilde{\mathbf{H}}_{t}^{S},\tilde{\mathbf{I}}_{t}].
    \label{equ6}
\end{equation}

\subsection{Temporal Module: LSTM}
To capture the temporal dependency of district-level features and infection cases, we adopt the existing LSTM model \cite{hochreiter1997long}. For $i \in \{1,2,...,n\}$, we extract the $i$-th row of matrices $\mathbf{H}_{t}^{A}, t\in [T-D_{1}+1,T]$ and feed them into an LSTM sequence (Figure~\ref{figure2}). Since we have already modelled the spatial dependency in the spatial module, in the temporal module we pass the node representations from $\mathbf{H}_{t}^{A}$ separately for different nodes, and these LSTM sequences share the identical structures and parameters.

\begin{figure}[h]
    \centering
    \includegraphics[width=1.0\linewidth]{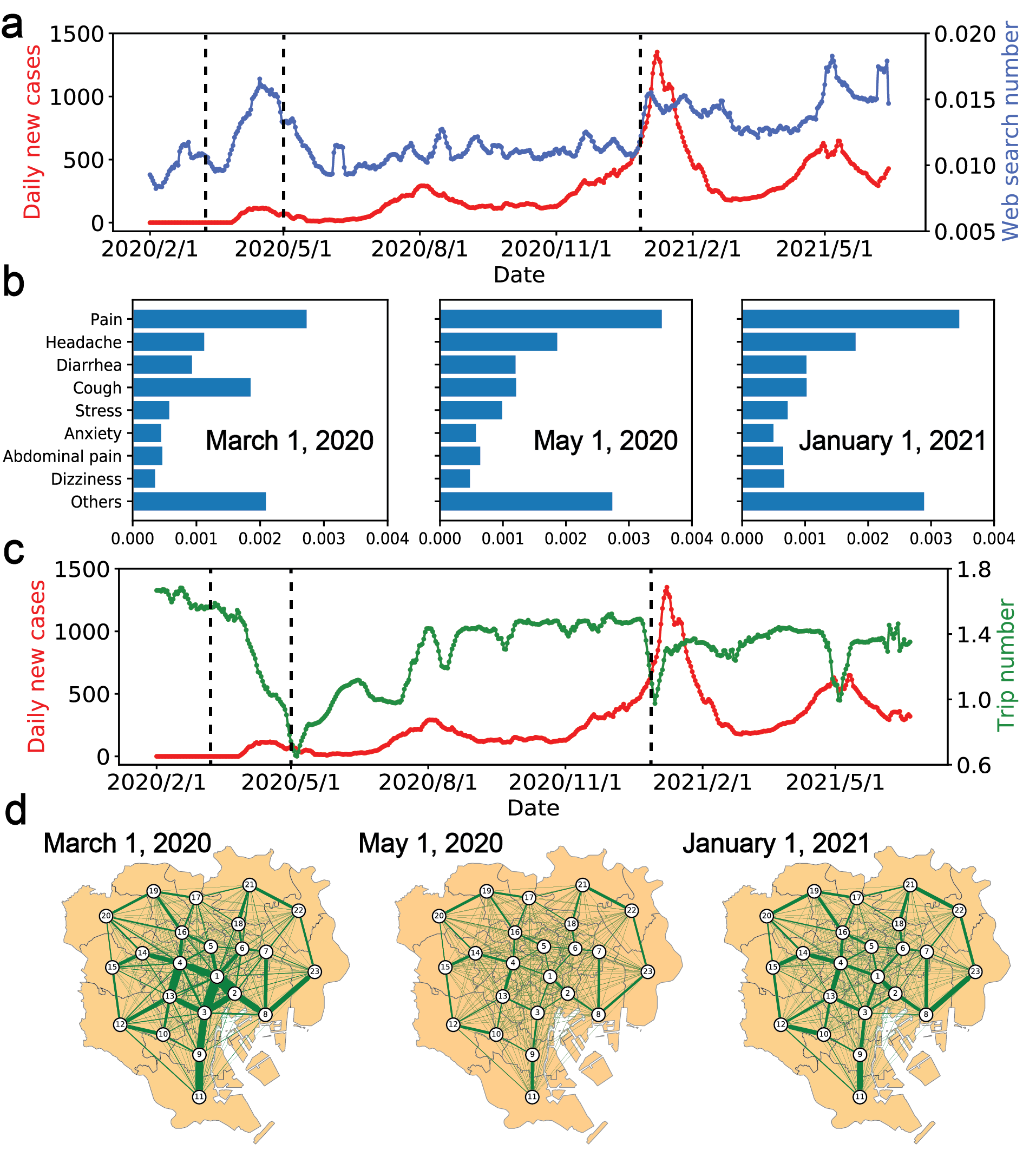}
    \vspace{-0.7cm}
    \caption{Statistics of mobility, web search frequency, and infection cases. (a) Dynamics of daily new cases and symptom-related web search frequency per user. (b) The distribution of web search frequency per user on three days. (c) Dynamics of daily new cases and inter-district trip number per user. (d) The number of inter-district trips on three days.}
    \label{figure3}
    \vspace{-0.4cm}
\end{figure}

\subsection{Loss Function}
Recall that our objective is to perform the infection case prediction for the next $D_{2}$ days, we define the loss function as the mean squared error, which is:
\begin{equation}
    \mathcal{L}_{T} = \frac{1}{D_{2}n}\sum_{t=T+1}^{T+D_{2}}\sum_{i=1}^{n}(I_{t,i}-\hat{I}_{t,i})^{2},
    \label{equ7}
\end{equation}
where $I_{t,i}$ and $\hat{I}_{t,i}$ denote the actual and predicted infection cases for the district $i$ on the day $t$. We present the training process as Algorithm in Supplement.

\section{Data}
\subsection{Mobility, Web Search, and Symptom Data}
We utilize four datasets: infection cases, mobile phone location data, web search data, and symptom data in Tokyo from Jan. 6, 2020, to May 15, 2021. The mobility and web search data were owned by Yahoo Japan Corporation with strict privacy protection regulations. The descriptions of these data are as follows:

\begin{itemize}
    \item \textbf{Infection data.} We access the daily new infection cases for 23 wards of Tokyo from the Tokyo COVID-19 Task Force website \cite{infection} which is maintained by Tokyo metropolitan government and open to public.
    \item \textbf{Mobility data.} Table \ref{table_11} in Supplement presents a sample piece of raw mobility data. The mobility data consists of a fake ID (after privacy protection operations), the user's geolocation in terms of longitude and latitude at a specific time. The data collection frequency was around 30 mins on average, which depends on the movements of the user: the frequency is higher if the user moves faster. In this study, the mobility data is utilized to identify Tokyo residents' home locations and mine the Mobility Matrix $\mathbf{E}_{t}$.

    \item \textbf{Web search data.} Table \ref{table_11} in Supplement presents a sample of raw web search record. Each record contains the user ID, web query words, as well as web search time. The web search data is used to get COVID-19 symptom-related web search counts. Note that Yahoo Japan Corporation maintains these data to promote users' experiences and does not share any individual-level data with other agencies.

    \item \textbf{Symptom data.} Based on two existing COVID-19 symptom studies \cite{sarker2020self,wang2021covid} and the COVID-19 symptom list released by the Centers for Disease Control and Prevention \cite{cdc}, we finally specify 44 symptoms, which are used to measure the counts of symptom-related web search records. These symptoms are shown in Table \ref{table_33} in Supplement. We translate these symptoms from English to Japanese, and then count the number of corresponding web searches in Japanese. Note that a few symptoms have a large number of web search counts, and we therefore utilize the most frequent $1 \leq n_{w}\leq 44$ symptoms in our prediction framework.  
\end{itemize}

\subsection{Data Preprocessing}
Note that the raw mobility and web search data are at the individual level. To prepare the features $\mathbf{E}_{t}$, $\mathbf{H}_{t}$ used in the machine-learning framework, we conduct the data preprocessing to aggregate the raw mobility and web search data to the urban district level. Please find the complete description of data prepossessing in the Appendix. 

The statistics of aggregate daily new infection cases, web search number, and inter-district trip number are shown in Figure~\ref{figure3}. The vertical dash lines in Figures~\ref{figure3}ac mark March 1, 2020, May 1, 2020, and January 1, 2021, whose web search distribution and mobility flow are visualized in Figures~\ref{figure3}bd, respectively. We find from Figure~\ref{figure3}a that the dynamic of symptom-related web search exhibits consistent peaks with daily infection cases, especially near April 2020, January 2021, and May 2021, which provides the direct evidence to inspire us to utilize the web search data in the multiwave pandemic prediction.


\section{RESULTS}
\subsection{Setting, Evaluation Metrics, Baselines}
We conduct experiments to predict two outbreak waves, i.e., \textbf{the third wave} (from Dec. 10, 2020 to  Feb. 7, 2021) and \textbf{the fourth wave} (from March 17 to May 15, 2021). The period for each experiment covers 10 months of observations where the train/validation/test ratio is 70\%/10\%/20\%. \textbf{We use the first 8 months (months 1 to 8) as training and validation data, and the last 2 months (months 9 and 10) as testing data. Note that the validation data has the size of 1 month and is evenly distributed from months 7 to 8.} Since most existing disease prediction studies \cite{panagopoulos2021transfer,rodriguez2021steering,wang2020examining} predict the infection at the week-level, we design three scenarios: $(D_{1},D_{2})=(21,7),(21,14),(21,21)$ (recall that we use the past $D_{1}$ days' features to predict the future $D_{2}$ days' infection cases).

We train the model using PyTorch and Adam optimizer \cite{kingma2014adam}. We use the validation set to determine the learning rate as 1e-4, the epoch number, the batch size, and the dropout rate as 100, 8, and 0.50, respectively. The experiments run on an Intel Xeon w-2155 3.3 GHz CPU and 32 GB of RAM. Root Mean Square Error (RMSE) and Mean Absolute Error (MAE) are utilized to evaluate the prediction performance: RMSE = $\sqrt{\frac{1}{D_{2}n}\sum_{t=T+1}^{T+D_{2}}\sum_{i=1}^{n}(I_{t,i}-\hat{I}_{t,i})^{2}}$,
MAE = $\frac{1}{D_{2}n}\sum_{t=T+1}^{T+D_{2}}\sum_{i=1}^{n}\vert I_{t,i}-\hat{I}_{t,i} \vert$.

To benchmark our model, we also implement 9 existing prediction models: (1) Historical average (HA) infection cases until the day $T$; (2) Historical average of the last $D_{1}$ days; (3,4) LSTM; (5,6) Seq2seq: encode the input infection case and decode the sequence using two separate LSTMs \cite{cho2014learning}; (7) ST-GNN: an Spatio-Temporal GNN model whose inputs are past infection cases and mobility patterns \cite{kapoor2020examining}; (8) MPNN+LSTM: a message passing neural network with the LSTM \cite{panagopoulos2021transfer}; (9) GraphLSTM: a prediction frameworks integrating GraphSage and LSTM \cite{sesti2021integrating}.

\subsection{Comparisons with Baseline Models}
\vspace{-0.2cm}
 \begin{figure}[H]
    \centering
    \includegraphics[width=1.0\linewidth]{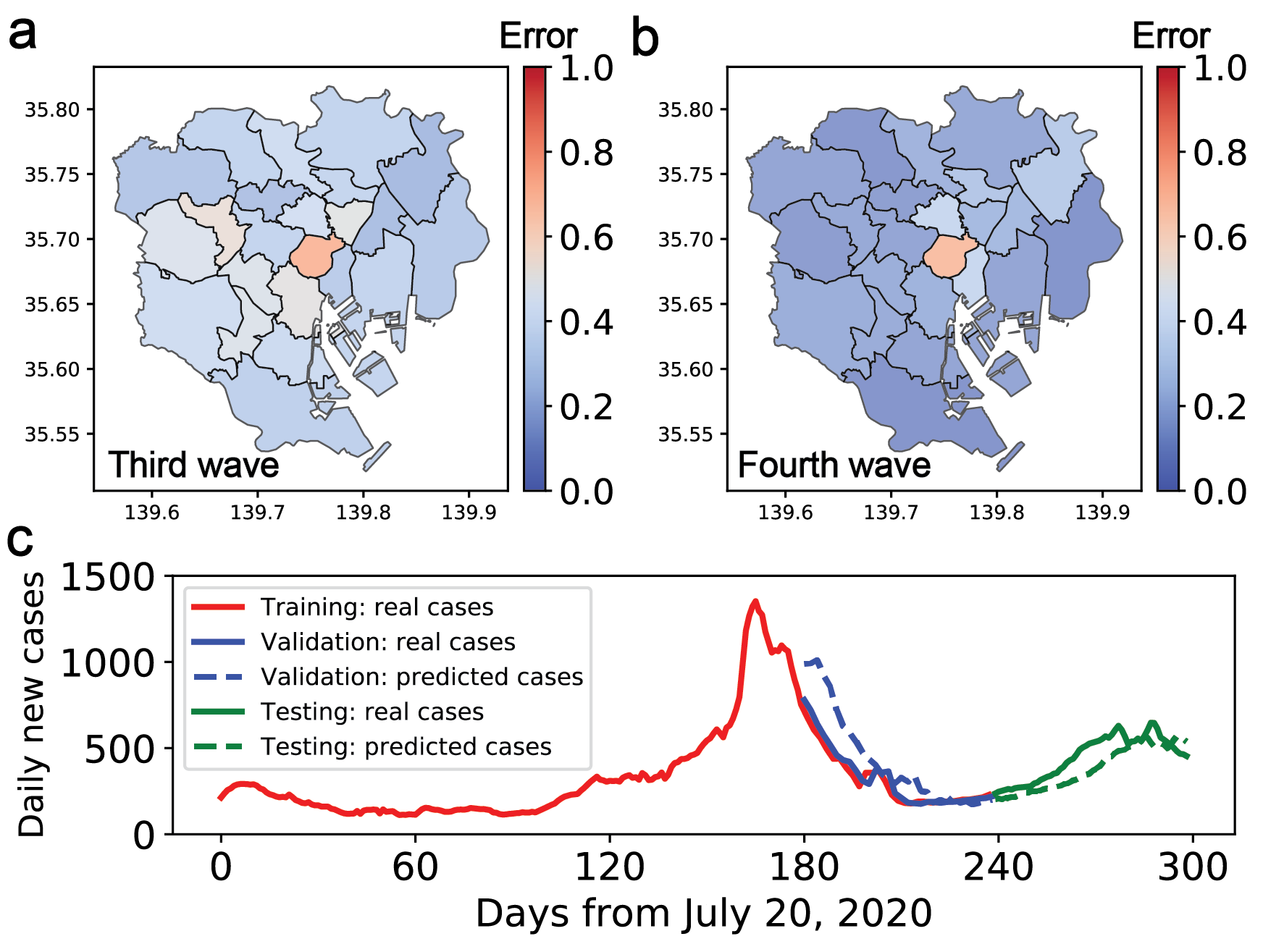}
    \vspace{-0.6cm}
    \caption{Prediction performance of the SAB-GNN. (a,b) The relative prediction error (i.e., $|I_{real}-I_{predict}|/|I_{real}|$) during the third wave (a) and the fourth wave (b).
    (c) Predicted cases and real cases for the fourth wave when $(D_{1},D_{2})=(21,7)$. Recall that days 180-240 are evenly divided into training data and validation data, so that train/validation/test =70\%/10\%/20\%.}
    \label{cross_1}
    \vspace{-0.3cm}
\end{figure}
We present the prediction performance of the SAB-GNN and other baseline models for the third wave and the fourth wave (Table \ref{table1}). Our proposed SAB-GNN model yields the smallest MAE and RMSE for most scenarios (5 out 6), which demonstrates our model captures the relationships between past infection, web search, mobility, and future infection. Comparing SAB-GNN and SAB-GNN-wsa (\underline{w}ithout \underline{s}ocial \underline{a}wareness recovery), the social awareness recovery mechanism decreases the prediction error for most cases (5 out of 6), and the web search frequency has higher predictability after the awareness recovery operation. 

\begin{table*}[h]
\centering
\begin{tabular}{cccccccccccccc}
\toprule[1.5pt]
\addlinespace[1.0pt]
            &          & \multicolumn{6}{c}{($D_{1},D_{2}$) for the third wave}                                                     & \multicolumn{6}{c}{($D_{1},D_{2}$) for the fourth wave}                                                    \\ \cline{3-14} 
Model       & Features & \multicolumn{2}{c}{(21,7)} & \multicolumn{2}{c}{(21,14)} & \multicolumn{2}{c}{(21,21)} & \multicolumn{2}{c}{(21,7)} & \multicolumn{2}{c}{(21,14)} & \multicolumn{2}{c}{(21,21)} \\
\addlinespace[-2pt]
\cmidrule(r){3-8} \cmidrule(l){9-14} 
\addlinespace[-2pt]
            &          & MAE          & RMSE        & MAE          & RMSE         & MAE          & RMSE         & MAE         & RMSE         & MAE          & RMSE         & MAE          & RMSE         \\ \hline
HA (all)    & I        & 22.63        & 26.45       & 22.53        & 27.22        & 22.16        & 27.65        & 8.76        & 10.04        & 9.14         & 10.59        & 9.41         & 11.08        \\
HA (X days) & I        & 14.15        & 17.02       & 16.92        & 20.72        & 19.14        & 23.81        & 4.52        & 5.76         & 5.52         & 7.02         & 6.47         & 8.23         \\
LSTM        & I        & 19.67        & 24.07       & 17.23        & 21.93        & 20.61        & 26.61        & 5.94        & 7.67         & 8.20          & 10.64        & 8.32         & 10.95        \\
LSTM        & I/W      & 20.61        & 24.89       & 20.72        & 25.91        & 19.93        & 25.94        & 7.43        & 9.79         & 8.28         & 10.55        & 8.11         & 10.63        \\
Seq2seq     & I        & 17.27        & 22.24       & 23.04        & 28.73        & 16.79        & 22.77        & 6.37        & 8.38         & 8.84         & 11.42        & 8.45         & 11.08        \\
Seq2seq     & I/W      & 15.69        & 19.83       & 21.76        & 27.46        & 19.66        & 25.69        & 5.96        & 7.74         & 7.93         & 9.86         & 10.43        & 12.94        \\
ST-GNN \cite{kapoor2020examining}      & I/W/M    & 20.20         & 25.86       & 21.53        & 27.32        & 20.24        & 26.36        & 5.46        & 6.95         & 7.63         & 9.97         & 10.54        & 13.96        \\
MPNN+LSTM \cite{panagopoulos2021transfer}  & I/W/M    & 12.40         & 17.31       & 16.30         & 21.73        & 19.78        & 25.59        & 3.34        & 4.49         & 5.27         & 7.97         & \textbf{4.80}          & \textbf{6.56}         \\
GraphLSTM \cite{sesti2021integrating}   & I/W/M    & 10.22            & 12.82           & 13.27            & 16.91            & 15.56            & 20.07            & 3.44           & 4.61            & 4.38            & 5.88            & 5.32            & 7.13           \\ \hline
SAB-GNN-wsa & I/W/M    & 10.75        & 13.43       & 12.72        & 16.33        & 15.61        & 20.10         & 3.32        & 4.46         & 4.63         & 6.18         & 5.03         & 6.77         \\
SAB-GNN     & I/W/M    & \textbf{8.03}         & \textbf{10.43}       & \textbf{11.23}        & \textbf{14.78}        & \textbf{13.76}        & \textbf{18.24}        & \textbf{3.25}        & \textbf{4.24}         & \textbf{4.28}         & \textbf{5.57}         & 5.25         & 6.82         \\ 
\addlinespace[-0.5pt]
\bottomrule[1.5pt]
\end{tabular}
\caption{\label{table1}Performance evaluation using past $D_{1}$ days' features to predict next $D_{2}$ days' infection ($I$: infection data; $W$: web search data; $M$: inter-district mobility data; SAB-GNN-wsa: the SAB-GNN model without the social awareness recovery).}
\vspace{-0.4cm}
\end{table*}

\begin{figure*}[ht]
    \centering
    \includegraphics[width=0.92\linewidth]{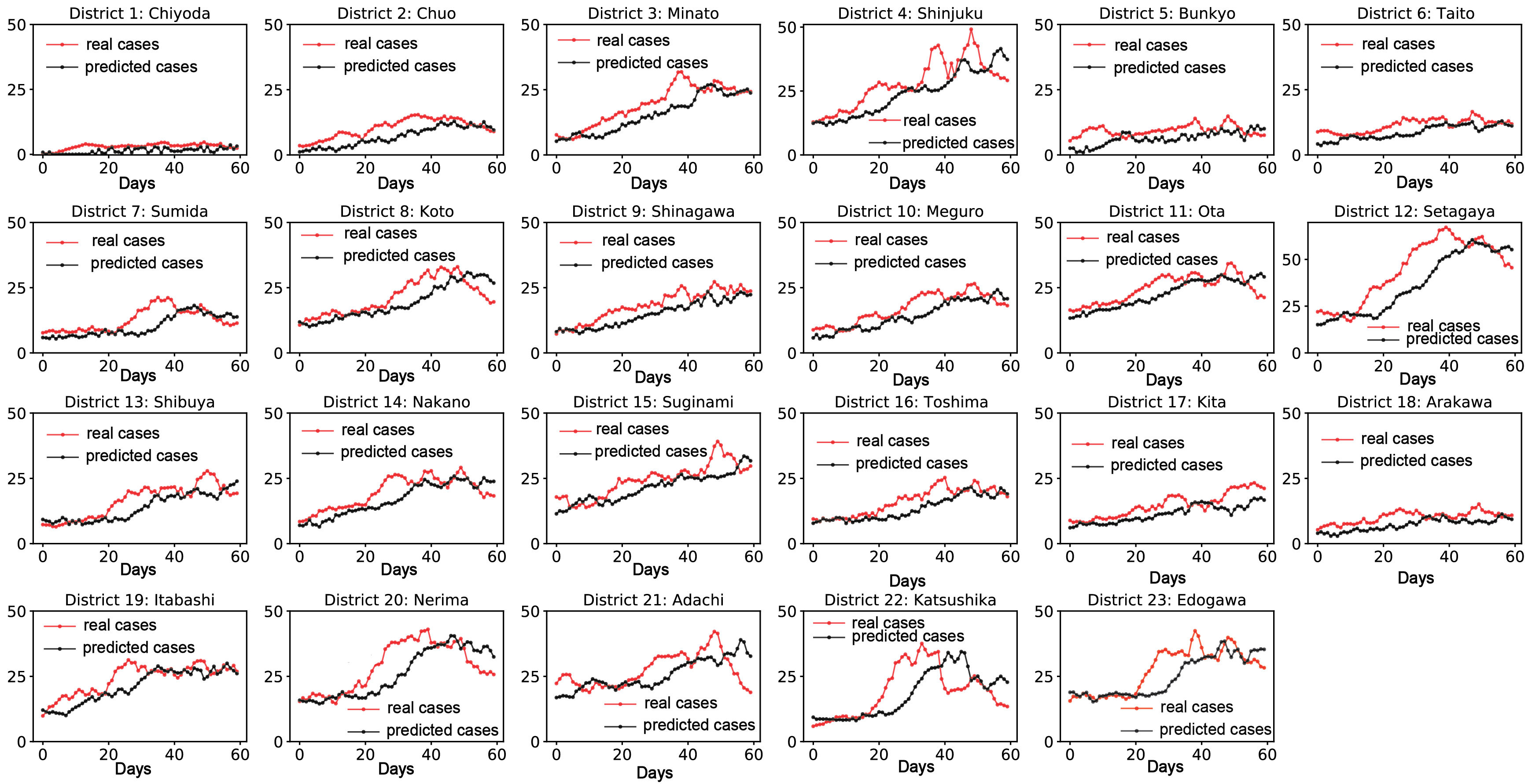}
    \vspace{-0.25cm}
    \caption{Predicted and real cases on different urban districts of SAB-GNN for the fourth wave when $(D_{1},D_{2})=(21,7)$.}
    \label{cross_2}
    \vspace{-0.25cm}
\end{figure*}
Besides, results from all models reveal a general tendency that the web search and mobility features result in prediction performance improvement, even if the past infection case is already a strong feature to connect with the future infection case. Finally, we observe from the LSTM and Seq2seq that simply concatenating the web search frequency embedding with the past infection cases is unable to consistently boost the prediction, which affirms the necessity of designing a suitable model architecture to utilize the web search data. For each model in Table 1, the running time for each scenario is within 20 minutes on the Ubuntu system with 32 GB RAM and 3.3 GHz Xeon w-2155 CPU.  
\subsection{Prediction Results for Different Urban Districts}
We visualize the prediction errors and compare the predicted cases with the actual cases for each urban district in Figures~\ref{cross_1}, \ref{cross_2}. As shown in Figures~\ref{cross_1}ab, the relative prediction errors for most districts are homogeneously lower than 0.50 except for the central district (i.e., Chiyoda). This reveals that SAB-GNN mines the relationships between features and future infection cases in a global manner thanks to the message passing mechanism among neighborhood nodes. The reason for the relatively large prediction error in Chiyoda (i.e., District 1) is because Chiyoda is where Imperial Palace locates and has quite a few real infection cases (Figure~\ref{cross_2}, the top-left panel). Figure~\ref{cross_1}c displays that the predicted daily case curve is able to capture the increasing tendency of actual cases from day 240 to day 280 and also maintains a small prediction error, which further demonstrates the power of our proposed SAB-GNN.

Figure~\ref{cross_2} exhibits that for most urban districts, the model is able to anticipate the general increasing tendency of infection cases during the fourth wave. This information is especially valuable for local public agencies to make timely preparations at the beginning of a wave. Note that the prediction of District 4 (i.e., Shinjuku) is not as accurate as other districts. One potential interpretation is that Shinjuku is a commercial area with many entertainment industries and thus does not have similar infection patterns as other districts. 

\subsection{Parameter Sensitivity}
\begin{figure}[t]
    \centering
    \includegraphics[width=0.82\linewidth]{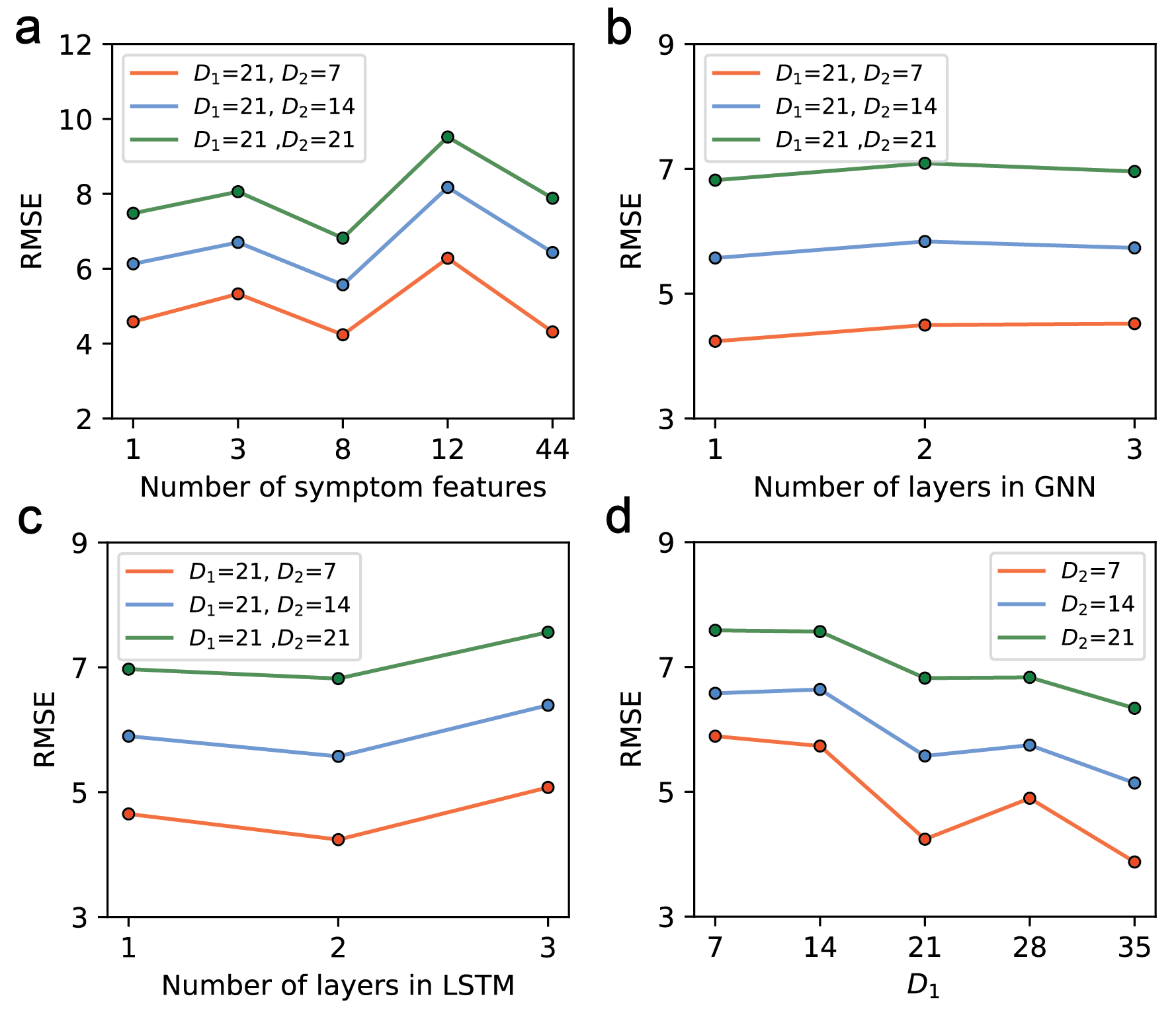}
    \vspace{-0.3cm}
    \caption{Parameter sensitivity study.}
    \label{figure5}
    \vspace{-0.3cm}
\end{figure}
\begin{figure}[h]
    \centering
    \includegraphics[width=0.9\linewidth]{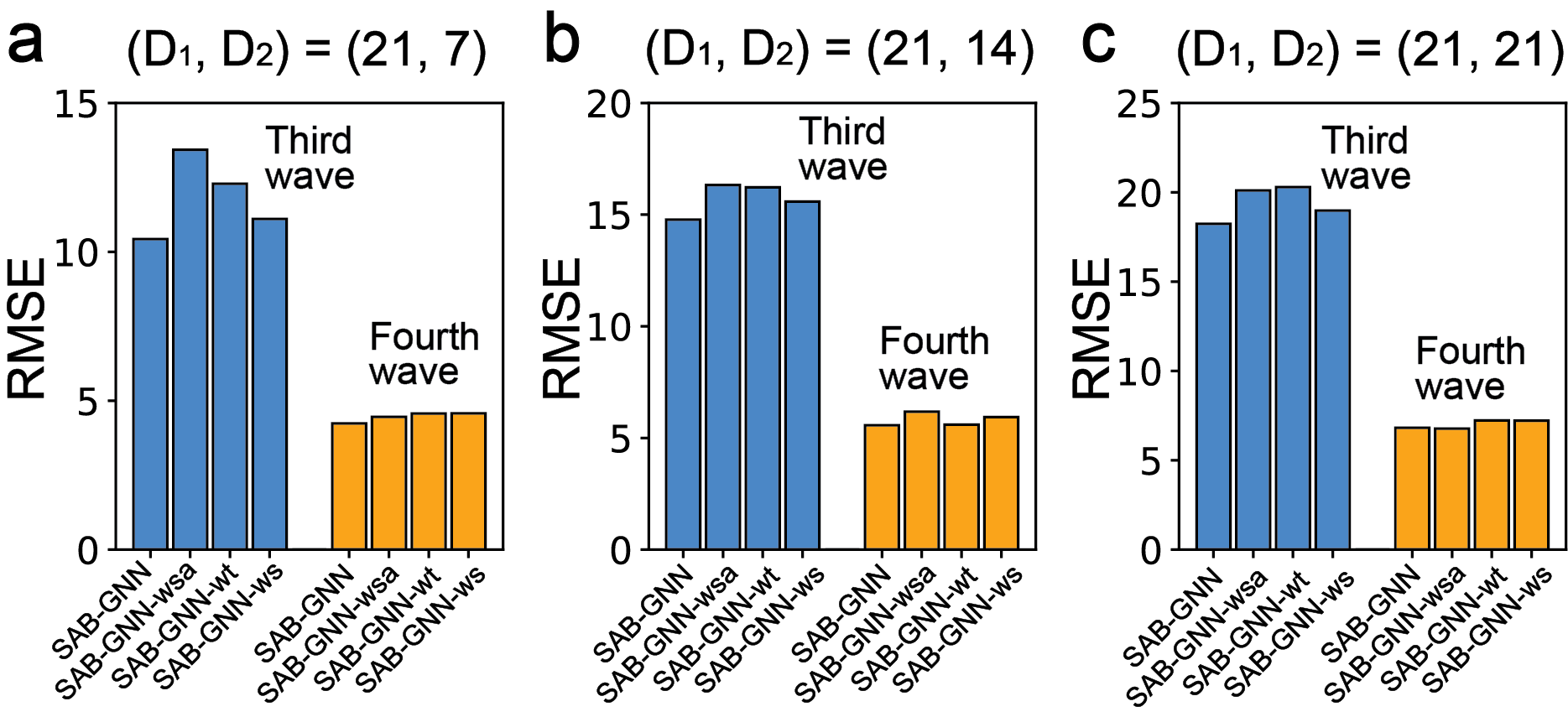}
    \vspace{-0.3cm}
    \caption{Ablation experiments. We implement the algorithms of SAB-GCN, SAB-GCN-wsa (without social awareness recovery module), SAB-GCN-wt (without temporal module), SAB-GCN-ws (without spatial module) for the third wave (the blue bar) and the fourth wave (the orange bar).}
    \label{figure6}
    \vspace{-0.5cm}
\end{figure}
The influences of model parameters in the SAB-GNN on the prediction performance are shown in Figure~\ref{figure5}. Figure~\ref{figure5}a displays the changes of RMSE for the fourth peak prediction with respect to $n_{w}$ (i.e., numbers of symptom-related web search features used in the model). Recall that there is a huge frequency discrepancy between different symptoms and we only feed the most frequent $n_{w}$ symptoms into our model. The result suggests that $n_w=8$ provides the best prediction performance. The underlying reason is that: when $n_{w}$ is small, more symptoms contribute stronger connections with future infection cases; when $n_{w}$ is large, many weakly-related symptoms bring noises to the training and thus harm the prediction performance. 

Next, we tune the numbers of layers $L_{1}, L_{2}$ in the GNN and LSTM modules (Figures~\ref{figure5}bc) and conclude that $(L_{1},L_{2})=(1,2)$ yields the best prediction. It is reasonable that the simple model with a small number of parameters is preferred in this infection case prediction task whose training data is at the hundreds level \cite{qiu2021miniseg} (each day is associated with one training sample). Finally, we test the model in Figure~\ref{figure5}d with varying $D_{1}$ and $D_{2}$ and observe the overall tendency that using more days' (i.e., $D_{1}$ is large) information to predict later fewer days' (i.e., $D_{2}$ is small) results in higher accuracy, which is consistent with the finding from the existing study \cite{panagopoulos2021transfer}. 
\subsection{Ablation Study}
To confirm the effect of each module in the SAB-GNN on prediction results, we perform the ablation experiments for both the third wave and fourth wave (Figure~\ref{figure6}). We remove one of the three modules in the SAB-GNN, and perform the prediction for the same temporal horizons as the full SAB-GNN model. We show quantitatively that the SAB-GNN model obtains lower prediction RMSE than other incomplete models, especially for the third wave, which reveals that both the spatial module, temporal module, and social awareness module contribute to the final prediction results in a positive manner. While existing studies have paid much attention to the spatial and temporal relationships between features and infection cases, we recommend introducing suitable social knowledge into the prediction, given the positive effect of the social awareness recovery mechanism.  
\section{DISCUSSION}
\textbf{Model performance analysis:} Underlying reasons for the superior prediction of SAB-GNN are threefold: (1) SAB-GNN leverages the power of GNN and LSTM to capture the spatio-temporal dynamic of disease spreading; (2) Web search features provide extra information as unconfirmed symptoms; (3) Social memory decay module properly reflects the social awareness decay as people get accustomed to COVID-19 during the pandemic. Note that we do not quantitatively test epidemiological models (e.g., SIR, SEIR) for two reasons: (1) Vanilla epidemiological models provide a good fit to the single wave or the first wave, but are insufficient to describe the multiwave infection outbreaks; (2) Quantitative experiments have demonstrated the predictability power of SEIR model is weaker than neural network model greatly \cite{schwabe2021predicting}.\\ \textbf{Policy implications:} This study demonstrates that web search data serves as a qualified pandemic surveillance indicator. It is able to complement existing mobility-based prediction models in various prediction platforms such as COVID-19 Forecast Hub \cite{Cramer2021-hub-dataset}, COVID-19 Mobility Impact Dashboard \cite{chang2021supporting}. Besides, policymakers can apply our tool on regional human text data collected from social media agencies (such as Twitter, Baidu, and Weibo) to forecast the next disease outbreak and make timely medical and social decisions, as part of the long-term battle against COVID-19. \\  
\textbf{Limitations:} We acknowledge two limitations in this study. First, we cannot ascertain used mobility and web search data from Yahoo! Japan evenly cover the Tokyo population from young children to old people. Extra work about representative of mobility and web search data in the whole population is needed before real-world deployment. Second, we assume the exponential decay of social awareness which lacks rigorous empirical and theoretical evidence. Future work can build a more solid decay function starting from the classical Collective Memory Theory, which describes the relationship between information (e.g., symptom-related web search) and human identity (e.g., infected, uninfected) from the field of cognition science \cite{roediger2015collective}.

\section{CONCLUSION} 
Motivated by the multiwave outbreak of COVID-19 across the globe, we establish the SAB-GNN to predict future infection cases. Except for the historical infection and mobility data, our approach utilizes the novel symptom-related web search data which provides alternative evidence of future waves. More importantly, we consider the social awareness decay effect and propose the social awareness recovery module to estimate the actual infection risks. Experiments on the third and fourth peaks of Tokyo affirm that the SAB-GNN outperforms other baseline models and captures the increasing trend of pandemic waves. Our method is applicable to many countries given the wide coverage of web search data. 

\section*{ACKNOWLEDGEMENTS}
We thank Yahoo Japan Corporation for the collaboration. Jiawei Xue and Satish Ukkusuri are partly supported by the National Science Foundation (Award Number: 1638311) grant for which the authors are grateful.


\bibliographystyle{ACM-Reference-Format}
\bibliography{base}


\begin{thebibliography}{61}


\ifx \showCODEN    \undefined \def \showCODEN     #1{\unskip}     \fi
\ifx \showDOI      \undefined \def \showDOI       #1{#1}\fi
\ifx \showISBNx    \undefined \def \showISBNx     #1{\unskip}     \fi
\ifx \showISBNxiii \undefined \def \showISBNxiii  #1{\unskip}     \fi
\ifx \showISSN     \undefined \def \showISSN      #1{\unskip}     \fi
\ifx \showLCCN     \undefined \def \showLCCN      #1{\unskip}     \fi
\ifx \shownote     \undefined \def \shownote      #1{#1}          \fi
\ifx \showarticletitle \undefined \def \showarticletitle #1{#1}   \fi
\ifx \showURL      \undefined \def \showURL       {\relax}        \fi
\providecommand\bibfield[2]{#2}
\providecommand\bibinfo[2]{#2}
\providecommand\natexlab[1]{#1}
\providecommand\showeprint[2][]{arXiv:#2}

\bibitem[\protect\citeauthoryear{Adiga, Wang, Hurt, Peddireddy, Porebski,
  Venkatramanan, Lewis, and Marathe}{Adiga et~al\mbox{.}}{2021}]%
        {adiga2021all}
\bibfield{author}{\bibinfo{person}{Aniruddha Adiga}, \bibinfo{person}{Lijing
  Wang}, \bibinfo{person}{Benjamin Hurt}, \bibinfo{person}{Akhil Peddireddy},
  \bibinfo{person}{Przemyslaw Porebski}, \bibinfo{person}{Srinivasan
  Venkatramanan}, \bibinfo{person}{Bryan~Leroy Lewis}, {and}
  \bibinfo{person}{Madhav Marathe}.} \bibinfo{year}{2021}\natexlab{}.
\newblock \showarticletitle{All models are useful: Bayesian ensembling for
  robust high resolution covid-19 forecasting}. In
  \bibinfo{booktitle}{\emph{Proceedings of the 27th ACM SIGKDD Conference on
  Knowledge Discovery \& Data Mining}}. \bibinfo{pages}{2505--2513}.
\newblock


\bibitem[\protect\citeauthoryear{Badr and Gardner}{Badr and Gardner}{2021}]%
        {badr2021limitations}
\bibfield{author}{\bibinfo{person}{Hamada~S Badr} {and}
  \bibinfo{person}{Lauren~M Gardner}.} \bibinfo{year}{2021}\natexlab{}.
\newblock \showarticletitle{Limitations of using mobile phone data to model
  COVID-19 transmission in the USA}.
\newblock \bibinfo{journal}{\emph{The Lancet Infectious Diseases}}
  \bibinfo{volume}{21}, \bibinfo{number}{5} (\bibinfo{year}{2021}),
  \bibinfo{pages}{e113}.
\newblock


\bibitem[\protect\citeauthoryear{Bastings, Titov, Aziz, Marcheggiani, and
  Sima'an}{Bastings et~al\mbox{.}}{2017}]%
        {bastings2017graph}
\bibfield{author}{\bibinfo{person}{Joost Bastings}, \bibinfo{person}{Ivan
  Titov}, \bibinfo{person}{Wilker Aziz}, \bibinfo{person}{Diego Marcheggiani},
  {and} \bibinfo{person}{Khalil Sima'an}.} \bibinfo{year}{2017}\natexlab{}.
\newblock \showarticletitle{Graph Convolutional Encoders for Syntax-aware
  Neural Machine Translation}. In \bibinfo{booktitle}{\emph{EMNLP}}.
\newblock


\bibitem[\protect\citeauthoryear{{Centers for Disease Control and
  Prevention}}{{Centers for Disease Control and Prevention}}{2021}]%
        {cdc}
\bibfield{author}{\bibinfo{person}{{Centers for Disease Control and
  Prevention}}.} \bibinfo{year}{2021}\natexlab{}.
\newblock \bibinfo{title}{Symptoms of COVID-19}.
\newblock \bibinfo{howpublished}{\url{
  https://www.cdc.gov/coronavirus/2019-ncov/symptoms-testing/symptoms.html}}.
\newblock


\bibitem[\protect\citeauthoryear{Chang, Pierson, Koh, Gerardin, Redbird,
  Grusky, and Leskovec}{Chang et~al\mbox{.}}{2020}]%
        {chang2020mobility}
\bibfield{author}{\bibinfo{person}{Serina Chang}, \bibinfo{person}{Emma
  Pierson}, \bibinfo{person}{Pang~Wei Koh}, \bibinfo{person}{Jaline Gerardin},
  \bibinfo{person}{Beth Redbird}, \bibinfo{person}{David Grusky}, {and}
  \bibinfo{person}{Jure Leskovec}.} \bibinfo{year}{2020}\natexlab{}.
\newblock \showarticletitle{Mobility network models of COVID-19 explain
  inequities and inform reopening}.
\newblock \bibinfo{journal}{\emph{Nature}} (\bibinfo{year}{2020}),
  \bibinfo{pages}{1--6}.
\newblock


\bibitem[\protect\citeauthoryear{Chang, Wilson, Lewis, Mehrab, Dudakiya,
  Pierson, Koh, Gerardin, Redbird, Grusky, et~al\mbox{.}}{Chang
  et~al\mbox{.}}{2021}]%
        {chang2021supporting}
\bibfield{author}{\bibinfo{person}{Serina Chang}, \bibinfo{person}{Mandy~L
  Wilson}, \bibinfo{person}{Bryan Lewis}, \bibinfo{person}{Zakaria Mehrab},
  \bibinfo{person}{Komal~K Dudakiya}, \bibinfo{person}{Emma Pierson},
  \bibinfo{person}{Pang~Wei Koh}, \bibinfo{person}{Jaline Gerardin},
  \bibinfo{person}{Beth Redbird}, \bibinfo{person}{David Grusky},
  {et~al\mbox{.}}} \bibinfo{year}{2021}\natexlab{}.
\newblock \showarticletitle{Supporting covid-19 policy response with
  large-scale mobility-based modeling}. In \bibinfo{booktitle}{\emph{SIGKDD}}.
  \bibinfo{pages}{2632--2642}.
\newblock


\bibitem[\protect\citeauthoryear{Chen, Li, Teo, Zou, Wang, Wang, and Zeng}{Chen
  et~al\mbox{.}}{2019}]%
        {chen2019gated}
\bibfield{author}{\bibinfo{person}{Cen Chen}, \bibinfo{person}{Kenli Li},
  \bibinfo{person}{Sin~G Teo}, \bibinfo{person}{Xiaofeng Zou},
  \bibinfo{person}{Kang Wang}, \bibinfo{person}{Jie Wang}, {and}
  \bibinfo{person}{Zeng Zeng}.} \bibinfo{year}{2019}\natexlab{}.
\newblock \showarticletitle{Gated residual recurrent graph neural networks for
  traffic prediction}. In \bibinfo{booktitle}{\emph{Proceedings of the AAAI
  conference on artificial intelligence}}, Vol.~\bibinfo{volume}{33}.
  \bibinfo{pages}{485--492}.
\newblock


\bibitem[\protect\citeauthoryear{Cheng}{Cheng}{1995}]%
        {cheng1995mean}
\bibfield{author}{\bibinfo{person}{Yizong Cheng}.}
  \bibinfo{year}{1995}\natexlab{}.
\newblock \showarticletitle{Mean shift, mode seeking, and clustering}.
\newblock \bibinfo{journal}{\emph{IEEE transactions on pattern analysis and
  machine intelligence}} \bibinfo{volume}{17}, \bibinfo{number}{8}
  (\bibinfo{year}{1995}), \bibinfo{pages}{790--799}.
\newblock


\bibitem[\protect\citeauthoryear{Chimmula and Zhang}{Chimmula and
  Zhang}{2020}]%
        {chimmula2020time}
\bibfield{author}{\bibinfo{person}{Vinay Kumar~Reddy Chimmula} {and}
  \bibinfo{person}{Lei Zhang}.} \bibinfo{year}{2020}\natexlab{}.
\newblock \showarticletitle{Time series forecasting of COVID-19 transmission in
  Canada using LSTM networks}.
\newblock \bibinfo{journal}{\emph{Chaos, Solitons \& Fractals}}
  (\bibinfo{year}{2020}), \bibinfo{pages}{109864}.
\newblock


\bibitem[\protect\citeauthoryear{Chinazzi}{Chinazzi}{2021}]%
        {usa_covid}
\bibfield{author}{\bibinfo{person}{Matteo Chinazzi}.} \bibinfo{year}{accessed
  August 2021}\natexlab{}.
\newblock \bibinfo{title}{Mobility, commuting, and contact patterns across the
  United States during the COVID-19 outbreak}.
\newblock
\newblock
\newblock
\shownote{Available online at \url{https://covid19.gleamproject.org/mobility}}.


\bibitem[\protect\citeauthoryear{Cho, Van~Merri{\"e}nboer, Gulcehre, Bahdanau,
  Bougares, Schwenk, and Bengio}{Cho et~al\mbox{.}}{2014}]%
        {cho2014learning}
\bibfield{author}{\bibinfo{person}{Kyunghyun Cho}, \bibinfo{person}{Bart
  Van~Merri{\"e}nboer}, \bibinfo{person}{Caglar Gulcehre},
  \bibinfo{person}{Dzmitry Bahdanau}, \bibinfo{person}{Fethi Bougares},
  \bibinfo{person}{Holger Schwenk}, {and} \bibinfo{person}{Yoshua Bengio}.}
  \bibinfo{year}{2014}\natexlab{}.
\newblock \showarticletitle{Learning phrase representations using RNN
  encoder-decoder for statistical machine translation}.
\newblock \bibinfo{journal}{\emph{arXiv preprint arXiv:1406.1078}}
  (\bibinfo{year}{2014}).
\newblock


\bibitem[\protect\citeauthoryear{Cramer~et al.}{Cramer~et al.}{2021}]%
        {Cramer2021-hub-dataset}
\bibfield{author}{\bibinfo{person}{Estee~Y Cramer~et al.}}
  \bibinfo{year}{2021}\natexlab{}.
\newblock \showarticletitle{The United States COVID-19 Forecast Hub dataset}.
\newblock \bibinfo{journal}{\emph{medRxiv}} (\bibinfo{year}{2021}).
\newblock
\urldef\tempurl%
\url{https://doi.org/10.1101/2021.11.04.21265886}
\showDOI{\tempurl}


\bibitem[\protect\citeauthoryear{Gao, Sharma, Qian, Glass, Spaeder, Romberg,
  Sun, and Xiao}{Gao et~al\mbox{.}}{2021}]%
        {gao2021stan}
\bibfield{author}{\bibinfo{person}{Junyi Gao}, \bibinfo{person}{Rakshith
  Sharma}, \bibinfo{person}{Cheng Qian}, \bibinfo{person}{Lucas~M Glass},
  \bibinfo{person}{Jeffrey Spaeder}, \bibinfo{person}{Justin Romberg},
  \bibinfo{person}{Jimeng Sun}, {and} \bibinfo{person}{Cao Xiao}.}
  \bibinfo{year}{2021}\natexlab{}.
\newblock \showarticletitle{STAN: spatio-temporal attention network for
  pandemic prediction using real-world evidence}.
\newblock \bibinfo{journal}{\emph{Journal of the American Medical Informatics
  Association}} \bibinfo{volume}{28}, \bibinfo{number}{4}
  (\bibinfo{year}{2021}), \bibinfo{pages}{733--743}.
\newblock


\bibitem[\protect\citeauthoryear{Goel, Hofman, Lahaie, Pennock, and Watts}{Goel
  et~al\mbox{.}}{2010}]%
        {goel2010predicting}
\bibfield{author}{\bibinfo{person}{Sharad Goel}, \bibinfo{person}{Jake~M
  Hofman}, \bibinfo{person}{S{\'e}bastien Lahaie}, \bibinfo{person}{David~M
  Pennock}, {and} \bibinfo{person}{Duncan~J Watts}.}
  \bibinfo{year}{2010}\natexlab{}.
\newblock \showarticletitle{Predicting consumer behavior with Web search}.
\newblock \bibinfo{journal}{\emph{Proceedings of the National academy of
  sciences}} \bibinfo{volume}{107}, \bibinfo{number}{41}
  (\bibinfo{year}{2010}), \bibinfo{pages}{17486--17490}.
\newblock


\bibitem[\protect\citeauthoryear{{Google Inc.}}{{Google Inc.}}{2021}]%
        {covid_google}
\bibfield{author}{\bibinfo{person}{{Google Inc.}}}
  \bibinfo{year}{2021}\natexlab{}.
\newblock \bibinfo{title}{Covid-19 infection cases}.
\newblock \bibinfo{howpublished}{\url{
  https://news.google.com/covid19/map?hl=en-US&gl=US&ceid=US\%3Aen}}.
\newblock


\bibitem[\protect\citeauthoryear{Hamilton, Ying, and Leskovec}{Hamilton
  et~al\mbox{.}}{2017}]%
        {hamilton2017inductive}
\bibfield{author}{\bibinfo{person}{Will Hamilton}, \bibinfo{person}{Zhitao
  Ying}, {and} \bibinfo{person}{Jure Leskovec}.}
  \bibinfo{year}{2017}\natexlab{}.
\newblock \showarticletitle{Inductive representation learning on large graphs}.
  In \bibinfo{booktitle}{\emph{Advances in neural information processing
  systems}}. \bibinfo{pages}{1024--1034}.
\newblock


\bibitem[\protect\citeauthoryear{Hisada, Murayama, Tsubouchi, Fujita, Yada,
  Wakamiya, and Aramaki}{Hisada et~al\mbox{.}}{2020}]%
        {hisada2020surveillance}
\bibfield{author}{\bibinfo{person}{Shohei Hisada}, \bibinfo{person}{Taichi
  Murayama}, \bibinfo{person}{Kota Tsubouchi}, \bibinfo{person}{Sumio Fujita},
  \bibinfo{person}{Shuntaro Yada}, \bibinfo{person}{Shoko Wakamiya}, {and}
  \bibinfo{person}{Eiji Aramaki}.} \bibinfo{year}{2020}\natexlab{}.
\newblock \showarticletitle{Surveillance of early stage COVID-19 clusters using
  search query logs and mobile device-based location information}.
\newblock \bibinfo{journal}{\emph{Scientific Reports}} \bibinfo{volume}{10},
  \bibinfo{number}{1} (\bibinfo{year}{2020}), \bibinfo{pages}{1--8}.
\newblock


\bibitem[\protect\citeauthoryear{Hochreiter and Schmidhuber}{Hochreiter and
  Schmidhuber}{1997}]%
        {hochreiter1997long}
\bibfield{author}{\bibinfo{person}{Sepp Hochreiter} {and}
  \bibinfo{person}{J{\"u}rgen Schmidhuber}.} \bibinfo{year}{1997}\natexlab{}.
\newblock \showarticletitle{Long short-term memory}.
\newblock \bibinfo{journal}{\emph{Neural computation}} \bibinfo{volume}{9},
  \bibinfo{number}{8} (\bibinfo{year}{1997}), \bibinfo{pages}{1735--1780}.
\newblock


\bibitem[\protect\citeauthoryear{Huang, Li, Jiang, Li, and Porter}{Huang
  et~al\mbox{.}}{2020}]%
        {huang2020twitter}
\bibfield{author}{\bibinfo{person}{Xiao Huang}, \bibinfo{person}{Zhenlong Li},
  \bibinfo{person}{Yuqin Jiang}, \bibinfo{person}{Xiaoming Li}, {and}
  \bibinfo{person}{Dwayne Porter}.} \bibinfo{year}{2020}\natexlab{}.
\newblock \showarticletitle{Twitter reveals human mobility dynamics during the
  COVID-19 pandemic}.
\newblock \bibinfo{journal}{\emph{PloS one}} \bibinfo{volume}{15},
  \bibinfo{number}{11} (\bibinfo{year}{2020}), \bibinfo{pages}{e0241957}.
\newblock


\bibitem[\protect\citeauthoryear{Jo, Kim, Huang, and Ni}{Jo
  et~al\mbox{.}}{2020}]%
        {jo2020condlstm}
\bibfield{author}{\bibinfo{person}{HyeongChan Jo}, \bibinfo{person}{Juhyun
  Kim}, \bibinfo{person}{Tzu-Chen Huang}, {and} \bibinfo{person}{Yu-Li Ni}.}
  \bibinfo{year}{2020}\natexlab{}.
\newblock \showarticletitle{condLSTM-Q: A novel deep learning model for
  predicting Covid-19 mortality in fine geographical Scale}.
\newblock \bibinfo{journal}{\emph{arXiv preprint arXiv:2011.11507}}
  (\bibinfo{year}{2020}).
\newblock


\bibitem[\protect\citeauthoryear{Ju, Song, Sun, Ye, Fan, Hou, Loparo, and
  Zhao}{Ju et~al\mbox{.}}{2021}]%
        {ju2021dr}
\bibfield{author}{\bibinfo{person}{Mingxuan Ju}, \bibinfo{person}{Wei Song},
  \bibinfo{person}{Shiyu Sun}, \bibinfo{person}{Yanfang Ye},
  \bibinfo{person}{Yujie Fan}, \bibinfo{person}{Shifu Hou},
  \bibinfo{person}{Kenneth Loparo}, {and} \bibinfo{person}{Liang Zhao}.}
  \bibinfo{year}{2021}\natexlab{}.
\newblock \showarticletitle{Dr. Emotion: Disentangled Representation Learning
  for Emotion Analysis on Social Media to Improve Community Resilience in the
  COVID-19 Era and Beyond}. In \bibinfo{booktitle}{\emph{Proceedings of the Web
  Conference 2021}}. \bibinfo{pages}{518--528}.
\newblock


\bibitem[\protect\citeauthoryear{Kang, Gao, Liang, Li, Rao, and Kruse}{Kang
  et~al\mbox{.}}{2020}]%
        {kang2020multiscale}
\bibfield{author}{\bibinfo{person}{Yuhao Kang}, \bibinfo{person}{Song Gao},
  \bibinfo{person}{Yunlei Liang}, \bibinfo{person}{Mingxiao Li},
  \bibinfo{person}{Jinmeng Rao}, {and} \bibinfo{person}{Jake Kruse}.}
  \bibinfo{year}{2020}\natexlab{}.
\newblock \showarticletitle{Multiscale dynamic human mobility flow dataset in
  the US during the COVID-19 epidemic}.
\newblock \bibinfo{journal}{\emph{Scientific data}} \bibinfo{volume}{7},
  \bibinfo{number}{1} (\bibinfo{year}{2020}), \bibinfo{pages}{1--13}.
\newblock


\bibitem[\protect\citeauthoryear{Kapoor, Ben, Liu, Perozzi, Barnes, Blais, and
  O'Banion}{Kapoor et~al\mbox{.}}{2020}]%
        {kapoor2020examining}
\bibfield{author}{\bibinfo{person}{Amol Kapoor}, \bibinfo{person}{Xue Ben},
  \bibinfo{person}{Luyang Liu}, \bibinfo{person}{Bryan Perozzi},
  \bibinfo{person}{Matt Barnes}, \bibinfo{person}{Martin Blais}, {and}
  \bibinfo{person}{Shawn O'Banion}.} \bibinfo{year}{2020}\natexlab{}.
\newblock \showarticletitle{Examining covid-19 forecasting using
  spatio-temporal graph neural networks}.
\newblock \bibinfo{journal}{\emph{arXiv preprint arXiv:2007.03113}}
  (\bibinfo{year}{2020}).
\newblock


\bibitem[\protect\citeauthoryear{Kargas, Qian, Sidiropoulos, Xiao, Glass, Sun,
  et~al\mbox{.}}{Kargas et~al\mbox{.}}{2021}]%
        {kargas2021stelar}
\bibfield{author}{\bibinfo{person}{Nikos Kargas}, \bibinfo{person}{Cheng Qian},
  \bibinfo{person}{Nicholas~D Sidiropoulos}, \bibinfo{person}{Cao Xiao},
  \bibinfo{person}{Lucas~M Glass}, \bibinfo{person}{Jimeng Sun},
  {et~al\mbox{.}}} \bibinfo{year}{2021}\natexlab{}.
\newblock \showarticletitle{STELAR: Spatio-temporal tensor factorization with
  latent epidemiological regularization}. In \bibinfo{booktitle}{\emph{35th
  AAAI Conference on Artificial Intelligence (AAAI)}}.
\newblock


\bibitem[\protect\citeauthoryear{Kingma and Ba}{Kingma and Ba}{2014}]%
        {kingma2014adam}
\bibfield{author}{\bibinfo{person}{Diederik~P Kingma} {and}
  \bibinfo{person}{Jimmy Ba}.} \bibinfo{year}{2014}\natexlab{}.
\newblock \showarticletitle{Adam: A method for stochastic optimization}.
\newblock \bibinfo{journal}{\emph{arXiv preprint arXiv:1412.6980}}
  (\bibinfo{year}{2014}).
\newblock


\bibitem[\protect\citeauthoryear{Kipf and Welling}{Kipf and Welling}{2016}]%
        {kipf2016semi}
\bibfield{author}{\bibinfo{person}{Thomas~N Kipf} {and} \bibinfo{person}{Max
  Welling}.} \bibinfo{year}{2016}\natexlab{}.
\newblock \showarticletitle{Semi-supervised classification with graph
  convolutional networks}.
\newblock \bibinfo{journal}{\emph{arXiv preprint arXiv:1609.02907}}
  (\bibinfo{year}{2016}).
\newblock


\bibitem[\protect\citeauthoryear{Kufel et~al\mbox{.}}{Kufel
  et~al\mbox{.}}{2020}]%
        {kufel2020arima}
\bibfield{author}{\bibinfo{person}{Tadeusz Kufel} {et~al\mbox{.}}}
  \bibinfo{year}{2020}\natexlab{}.
\newblock \showarticletitle{ARIMA-based forecasting of the dynamics of
  confirmed Covid-19 cases for selected European countries}.
\newblock \bibinfo{journal}{\emph{Equilibrium. Quarterly Journal of Economics
  and Economic Policy}} \bibinfo{volume}{15}, \bibinfo{number}{2}
  (\bibinfo{year}{2020}), \bibinfo{pages}{181--204}.
\newblock


\bibitem[\protect\citeauthoryear{Leung, Wu, Liu, and Leung}{Leung
  et~al\mbox{.}}{2020}]%
        {leung2020first}
\bibfield{author}{\bibinfo{person}{Kathy Leung}, \bibinfo{person}{Joseph~T Wu},
  \bibinfo{person}{Di Liu}, {and} \bibinfo{person}{Gabriel~M Leung}.}
  \bibinfo{year}{2020}\natexlab{}.
\newblock \showarticletitle{First-wave COVID-19 transmissibility and severity
  in China outside Hubei after control measures, and second-wave scenario
  planning: a modelling impact assessment}.
\newblock \bibinfo{journal}{\emph{The Lancet}} \bibinfo{volume}{395},
  \bibinfo{number}{10233} (\bibinfo{year}{2020}), \bibinfo{pages}{1382--1393}.
\newblock


\bibitem[\protect\citeauthoryear{Levin, Chao, Wenger, and Proctor}{Levin
  et~al\mbox{.}}{2021}]%
        {levin2021insights}
\bibfield{author}{\bibinfo{person}{Roman Levin}, \bibinfo{person}{Dennis~L
  Chao}, \bibinfo{person}{Edward~A Wenger}, {and} \bibinfo{person}{Joshua~L
  Proctor}.} \bibinfo{year}{2021}\natexlab{}.
\newblock \showarticletitle{Insights into population behavior during the
  COVID-19 pandemic from cell phone mobility data and manifold learning}.
\newblock \bibinfo{journal}{\emph{Nature Computational Science}}
  \bibinfo{volume}{1}, \bibinfo{number}{9} (\bibinfo{year}{2021}),
  \bibinfo{pages}{588--597}.
\newblock


\bibitem[\protect\citeauthoryear{Moghadas, Shoukat, Fitzpatrick, Wells, Sah,
  Pandey, Sachs, Wang, Meyers, Singer, et~al\mbox{.}}{Moghadas
  et~al\mbox{.}}{2020}]%
        {moghadas2020projecting}
\bibfield{author}{\bibinfo{person}{Seyed~M Moghadas}, \bibinfo{person}{Affan
  Shoukat}, \bibinfo{person}{Meagan~C Fitzpatrick}, \bibinfo{person}{Chad~R
  Wells}, \bibinfo{person}{Pratha Sah}, \bibinfo{person}{Abhishek Pandey},
  \bibinfo{person}{Jeffrey~D Sachs}, \bibinfo{person}{Zheng Wang},
  \bibinfo{person}{Lauren~A Meyers}, \bibinfo{person}{Burton~H Singer},
  {et~al\mbox{.}}} \bibinfo{year}{2020}\natexlab{}.
\newblock \showarticletitle{Projecting hospital utilization during the COVID-19
  outbreaks in the United States}.
\newblock \bibinfo{journal}{\emph{Proceedings of the National Academy of
  Sciences}} \bibinfo{volume}{117}, \bibinfo{number}{16}
  (\bibinfo{year}{2020}), \bibinfo{pages}{9122--9126}.
\newblock


\bibitem[\protect\citeauthoryear{Narasimhan, Lazebnik, and Schwing}{Narasimhan
  et~al\mbox{.}}{2018}]%
        {narasimhan2018out}
\bibfield{author}{\bibinfo{person}{Medhini Narasimhan},
  \bibinfo{person}{Svetlana Lazebnik}, {and} \bibinfo{person}{Alexander~G
  Schwing}.} \bibinfo{year}{2018}\natexlab{}.
\newblock \showarticletitle{Out of the box: Reasoning with graph convolution
  nets for factual visual question answering}.
\newblock \bibinfo{journal}{\emph{Advances in Neural Information Processing
  Systems}}  \bibinfo{volume}{2018} (\bibinfo{year}{2018}),
  \bibinfo{pages}{2654--2665}.
\newblock


\bibitem[\protect\citeauthoryear{Organization}{Organization}{2022}]%
        {2022who}
\bibfield{author}{\bibinfo{person}{World~Health Organization}.}
  \bibinfo{year}{accessed June 2022}\natexlab{}.
\newblock \bibinfo{title}{Coronavirus disease (COVID-19) pandemic}.
\newblock
\newblock
\newblock
\shownote{Available online at \url{https://covid19.who.int/table}}.


\bibitem[\protect\citeauthoryear{Panagopoulos, Nikolentzos, and
  Vazirgiannis}{Panagopoulos et~al\mbox{.}}{2021}]%
        {panagopoulos2021transfer}
\bibfield{author}{\bibinfo{person}{George Panagopoulos},
  \bibinfo{person}{Giannis Nikolentzos}, {and} \bibinfo{person}{Michalis
  Vazirgiannis}.} \bibinfo{year}{2021}\natexlab{}.
\newblock \showarticletitle{Transfer Graph Neural Networks for Pandemic
  Forecasting}. In \bibinfo{booktitle}{\emph{Proceedings of the AAAI Conference
  on Artificial Intelligence}}, Vol.~\bibinfo{volume}{35}.
  \bibinfo{pages}{4838--4845}.
\newblock


\bibitem[\protect\citeauthoryear{Peng, Wang, Du, Bhuiyan, Ma, Liu, Wang, Yang,
  Du, Wang, et~al\mbox{.}}{Peng et~al\mbox{.}}{2020}]%
        {peng2020spatial}
\bibfield{author}{\bibinfo{person}{Hao Peng}, \bibinfo{person}{Hongfei Wang},
  \bibinfo{person}{Bowen Du}, \bibinfo{person}{Md~Zakirul~Alam Bhuiyan},
  \bibinfo{person}{Hongyuan Ma}, \bibinfo{person}{Jianwei Liu},
  \bibinfo{person}{Lihong Wang}, \bibinfo{person}{Zeyu Yang},
  \bibinfo{person}{Linfeng Du}, \bibinfo{person}{Senzhang Wang},
  {et~al\mbox{.}}} \bibinfo{year}{2020}\natexlab{}.
\newblock \showarticletitle{Spatial temporal incidence dynamic graph neural
  networks for traffic flow forecasting}.
\newblock \bibinfo{journal}{\emph{Information Sciences}}  \bibinfo{volume}{521}
  (\bibinfo{year}{2020}), \bibinfo{pages}{277--290}.
\newblock


\bibitem[\protect\citeauthoryear{Qian, Sun, and Ukkusuri}{Qian
  et~al\mbox{.}}{2021}]%
        {qian2021scaling}
\bibfield{author}{\bibinfo{person}{Xinwu Qian}, \bibinfo{person}{Lijun Sun},
  {and} \bibinfo{person}{Satish~V Ukkusuri}.} \bibinfo{year}{2021}\natexlab{}.
\newblock \showarticletitle{Scaling of contact networks for epidemic spreading
  in urban transit systems}.
\newblock \bibinfo{journal}{\emph{Scientific reports}} \bibinfo{volume}{11},
  \bibinfo{number}{1} (\bibinfo{year}{2021}), \bibinfo{pages}{1--12}.
\newblock


\bibitem[\protect\citeauthoryear{Qian, Xue, and Ukkusuri}{Qian
  et~al\mbox{.}}{2020}]%
        {qian2020modeling}
\bibfield{author}{\bibinfo{person}{Xinwu Qian}, \bibinfo{person}{Jiawei Xue},
  {and} \bibinfo{person}{Satish~V Ukkusuri}.} \bibinfo{year}{2020}\natexlab{}.
\newblock \showarticletitle{Modeling disease spreading with adaptive behavior
  considering local and global information dissemination}.
\newblock \bibinfo{journal}{\emph{arXiv preprint arXiv:2008.10853}}
  (\bibinfo{year}{2020}).
\newblock


\bibitem[\protect\citeauthoryear{Qiu, Liu, Li, and Xu}{Qiu
  et~al\mbox{.}}{2021}]%
        {qiu2021miniseg}
\bibfield{author}{\bibinfo{person}{Yu Qiu}, \bibinfo{person}{Yun Liu},
  \bibinfo{person}{Shijie Li}, {and} \bibinfo{person}{Jing Xu}.}
  \bibinfo{year}{2021}\natexlab{}.
\newblock \showarticletitle{MiniSeg: An Extremely Minimum Network for Efficient
  COVID-19 Segmentation}. In \bibinfo{booktitle}{\emph{Proceedings of the AAAI
  Conference on Artificial Intelligence}}, Vol.~\bibinfo{volume}{35}.
  \bibinfo{pages}{4846--4854}.
\newblock


\bibitem[\protect\citeauthoryear{Rodriguez, Muralidhar, Adhikari, Tabassum,
  Ramakrishnan, and Prakash}{Rodriguez et~al\mbox{.}}{2021}]%
        {rodriguez2021steering}
\bibfield{author}{\bibinfo{person}{Alexander Rodriguez},
  \bibinfo{person}{Nikhil Muralidhar}, \bibinfo{person}{Bijaya Adhikari},
  \bibinfo{person}{Anika Tabassum}, \bibinfo{person}{Naren Ramakrishnan}, {and}
  \bibinfo{person}{B~Aditya Prakash}.} \bibinfo{year}{2021}\natexlab{}.
\newblock \showarticletitle{Steering a Historical Disease Forecasting Model
  Under a Pandemic: Case of Flu and COVID-19}. In
  \bibinfo{booktitle}{\emph{Proceedings of AAAI}}.
\newblock


\bibitem[\protect\citeauthoryear{Roediger~III and Abel}{Roediger~III and
  Abel}{2015}]%
        {roediger2015collective}
\bibfield{author}{\bibinfo{person}{Henry~L Roediger~III} {and}
  \bibinfo{person}{Magdalena Abel}.} \bibinfo{year}{2015}\natexlab{}.
\newblock \showarticletitle{Collective memory: a new arena of cognitive study}.
\newblock \bibinfo{journal}{\emph{Trends in cognitive sciences}}
  \bibinfo{volume}{19}, \bibinfo{number}{7} (\bibinfo{year}{2015}),
  \bibinfo{pages}{359--361}.
\newblock


\bibitem[\protect\citeauthoryear{Sarker, Lakamana, Hogg-Bremer, Xie, Al-Garadi,
  and Yang}{Sarker et~al\mbox{.}}{2020}]%
        {sarker2020self}
\bibfield{author}{\bibinfo{person}{Abeed Sarker}, \bibinfo{person}{Sahithi
  Lakamana}, \bibinfo{person}{Whitney Hogg-Bremer}, \bibinfo{person}{Angel
  Xie}, \bibinfo{person}{Mohammed~Ali Al-Garadi}, {and}
  \bibinfo{person}{Yuan-Chi Yang}.} \bibinfo{year}{2020}\natexlab{}.
\newblock \showarticletitle{Self-reported COVID-19 symptoms on Twitter: an
  analysis and a research resource}.
\newblock \bibinfo{journal}{\emph{Journal of the American Medical Informatics
  Association}} \bibinfo{volume}{27}, \bibinfo{number}{8}
  (\bibinfo{year}{2020}), \bibinfo{pages}{1310--1315}.
\newblock


\bibitem[\protect\citeauthoryear{Schwabe, Persson, and Feuerriegel}{Schwabe
  et~al\mbox{.}}{2021}]%
        {schwabe2021predicting}
\bibfield{author}{\bibinfo{person}{Amray Schwabe}, \bibinfo{person}{Joel
  Persson}, {and} \bibinfo{person}{Stefan Feuerriegel}.}
  \bibinfo{year}{2021}\natexlab{}.
\newblock \showarticletitle{Predicting COVID-19 Spread from Large-Scale
  Mobility Data}. In \bibinfo{booktitle}{\emph{KDD 2021}}.
\newblock


\bibitem[\protect\citeauthoryear{Sesti, Garau-Luis, Crawley, and Cameron}{Sesti
  et~al\mbox{.}}{2021}]%
        {sesti2021integrating}
\bibfield{author}{\bibinfo{person}{Nathan Sesti}, \bibinfo{person}{Juan~Jose
  Garau-Luis}, \bibinfo{person}{Edward Crawley}, {and} \bibinfo{person}{Bruce
  Cameron}.} \bibinfo{year}{2021}\natexlab{}.
\newblock \showarticletitle{Integrating LSTMs and GNNs for COVID-19
  Forecasting}.
\newblock \bibinfo{journal}{\emph{arXiv preprint arXiv:2108.10052}}
  (\bibinfo{year}{2021}).
\newblock


\bibitem[\protect\citeauthoryear{Shahid, Zameer, and Muneeb}{Shahid
  et~al\mbox{.}}{2020}]%
        {shahid2020predictions}
\bibfield{author}{\bibinfo{person}{Farah Shahid}, \bibinfo{person}{Aneela
  Zameer}, {and} \bibinfo{person}{Muhammad Muneeb}.}
  \bibinfo{year}{2020}\natexlab{}.
\newblock \showarticletitle{Predictions for COVID-19 with deep learning models
  of LSTM, GRU and Bi-LSTM}.
\newblock \bibinfo{journal}{\emph{Chaos, Solitons \& Fractals}}
  \bibinfo{volume}{140} (\bibinfo{year}{2020}), \bibinfo{pages}{110212}.
\newblock


\bibitem[\protect\citeauthoryear{{Tokyo COVID-19 Task Force Website}}{{Tokyo
  COVID-19 Task Force Website}}{2021}]%
        {infection}
\bibfield{author}{\bibinfo{person}{{Tokyo COVID-19 Task Force Website}}.}
  \bibinfo{year}{2021}\natexlab{}.
\newblock \bibinfo{title}{Covid-19 infection cases}.
\newblock
  \bibinfo{howpublished}{\url{https://github.com/codeforshinjuku/covid19/blob/master/dist/patient.json}}.
\newblock


\bibitem[\protect\citeauthoryear{Veli{\v{c}}kovi{\'c}, Cucurull, Casanova,
  Romero, Lio, and Bengio}{Veli{\v{c}}kovi{\'c} et~al\mbox{.}}{2017}]%
        {velivckovic2017graph}
\bibfield{author}{\bibinfo{person}{Petar Veli{\v{c}}kovi{\'c}},
  \bibinfo{person}{Guillem Cucurull}, \bibinfo{person}{Arantxa Casanova},
  \bibinfo{person}{Adriana Romero}, \bibinfo{person}{Pietro Lio}, {and}
  \bibinfo{person}{Yoshua Bengio}.} \bibinfo{year}{2017}\natexlab{}.
\newblock \showarticletitle{Graph attention networks}.
\newblock \bibinfo{journal}{\emph{arXiv preprint arXiv:1710.10903}}
  (\bibinfo{year}{2017}).
\newblock


\bibitem[\protect\citeauthoryear{Wang, Abu-el Rub, Gray, Pham, Zhou, Manion,
  Liu, Song, Xu, Rouhizadeh, et~al\mbox{.}}{Wang et~al\mbox{.}}{2021}]%
        {wang2021covid}
\bibfield{author}{\bibinfo{person}{Jingqi Wang}, \bibinfo{person}{Noor Abu-el
  Rub}, \bibinfo{person}{Josh Gray}, \bibinfo{person}{Huy~Anh Pham},
  \bibinfo{person}{Yujia Zhou}, \bibinfo{person}{Frank~J Manion},
  \bibinfo{person}{Mei Liu}, \bibinfo{person}{Xing Song}, \bibinfo{person}{Hua
  Xu}, \bibinfo{person}{Masoud Rouhizadeh}, {et~al\mbox{.}}}
  \bibinfo{year}{2021}\natexlab{}.
\newblock \showarticletitle{COVID-19 SignSym: a fast adaptation of a general
  clinical NLP tool to identify and normalize COVID-19 signs and symptoms to
  OMOP common data model}.
\newblock \bibinfo{journal}{\emph{Journal of the American Medical Informatics
  Association}} \bibinfo{volume}{28}, \bibinfo{number}{6}
  (\bibinfo{year}{2021}), \bibinfo{pages}{1275--1283}.
\newblock


\bibitem[\protect\citeauthoryear{Wang, Adiga, Venkatramanan, Chen, Lewis, and
  Marathe}{Wang et~al\mbox{.}}{2020}]%
        {wang2020examining}
\bibfield{author}{\bibinfo{person}{Lijing Wang}, \bibinfo{person}{Aniruddha
  Adiga}, \bibinfo{person}{Srinivasan Venkatramanan},
  \bibinfo{person}{Jiangzhuo Chen}, \bibinfo{person}{Bryan Lewis}, {and}
  \bibinfo{person}{Madhav Marathe}.} \bibinfo{year}{2020}\natexlab{}.
\newblock \showarticletitle{Examining Deep Learning Models with Multiple Data
  Sources for COVID-19 Forecasting}. In \bibinfo{booktitle}{\emph{2020 IEEE
  International Conference on Big Data (Big Data)}}. IEEE,
  \bibinfo{pages}{3846--3855}.
\newblock


\bibitem[\protect\citeauthoryear{Xiao, Zhou, Huang, Zhuo, Liu, Xiong, and
  Dou}{Xiao et~al\mbox{.}}{2021}]%
        {xiao2021c}
\bibfield{author}{\bibinfo{person}{Congxi Xiao}, \bibinfo{person}{Jingbo Zhou},
  \bibinfo{person}{Jizhou Huang}, \bibinfo{person}{An Zhuo},
  \bibinfo{person}{Ji Liu}, \bibinfo{person}{Haoyi Xiong}, {and}
  \bibinfo{person}{Dejing Dou}.} \bibinfo{year}{2021}\natexlab{}.
\newblock \showarticletitle{C-Watcher: A Framework for Early Detection of
  High-Risk Neighborhoods Ahead of COVID-19 Outbreak}. In
  \bibinfo{booktitle}{\emph{Proceedings of the AAAI Conference on Artificial
  Intelligence}}, Vol.~\bibinfo{volume}{35}. \bibinfo{pages}{4892--4900}.
\newblock


\bibitem[\protect\citeauthoryear{Xiong, Hu, Yang, Luo, and Zhang}{Xiong
  et~al\mbox{.}}{2020}]%
        {xiong2020mobile}
\bibfield{author}{\bibinfo{person}{Chenfeng Xiong}, \bibinfo{person}{Songhua
  Hu}, \bibinfo{person}{Mofeng Yang}, \bibinfo{person}{Weiyu Luo}, {and}
  \bibinfo{person}{Lei Zhang}.} \bibinfo{year}{2020}\natexlab{}.
\newblock \showarticletitle{Mobile device data reveal the dynamics in a
  positive relationship between human mobility and COVID-19 infections}.
\newblock \bibinfo{journal}{\emph{Proceedings of the National Academy of
  Sciences}} \bibinfo{volume}{117}, \bibinfo{number}{44}
  (\bibinfo{year}{2020}), \bibinfo{pages}{27087--27089}.
\newblock


\bibitem[\protect\citeauthoryear{Xue, Jiang, Liang, Pang, Yabe, Ukkusuri, and
  Ma}{Xue et~al\mbox{.}}{2022}]%
        {xue2022quantifying}
\bibfield{author}{\bibinfo{person}{Jiawei Xue}, \bibinfo{person}{Nan Jiang},
  \bibinfo{person}{Senwei Liang}, \bibinfo{person}{Qiyuan Pang},
  \bibinfo{person}{Takahiro Yabe}, \bibinfo{person}{Satish~V Ukkusuri}, {and}
  \bibinfo{person}{Jianzhu Ma}.} \bibinfo{year}{2022}\natexlab{}.
\newblock \showarticletitle{Quantifying the spatial homogeneity of urban road
  networks via graph neural networks}.
\newblock \bibinfo{journal}{\emph{Nature Machine Intelligence}}
  \bibinfo{volume}{4}, \bibinfo{number}{3} (\bibinfo{year}{2022}),
  \bibinfo{pages}{246--257}.
\newblock


\bibitem[\protect\citeauthoryear{Yabe, Tsubouchi, Fujiwara, Wada, Sekimoto, and
  Ukkusuri}{Yabe et~al\mbox{.}}{2020}]%
        {yabe2020non}
\bibfield{author}{\bibinfo{person}{Takahiro Yabe}, \bibinfo{person}{Kota
  Tsubouchi}, \bibinfo{person}{Naoya Fujiwara}, \bibinfo{person}{Takayuki
  Wada}, \bibinfo{person}{Yoshihide Sekimoto}, {and} \bibinfo{person}{Satish~V
  Ukkusuri}.} \bibinfo{year}{2020}\natexlab{}.
\newblock \showarticletitle{Non-compulsory measures sufficiently reduced human
  mobility in Tokyo during the COVID-19 epidemic}.
\newblock \bibinfo{journal}{\emph{Scientific reports}} \bibinfo{volume}{10},
  \bibinfo{number}{1} (\bibinfo{year}{2020}), \bibinfo{pages}{1--9}.
\newblock


\bibitem[\protect\citeauthoryear{Yabe, Tsubouchi, Sekimoto, and Ukkusuri}{Yabe
  et~al\mbox{.}}{2021}]%
        {yabe2021early}
\bibfield{author}{\bibinfo{person}{Takahiro Yabe}, \bibinfo{person}{Kota
  Tsubouchi}, \bibinfo{person}{Yoshihide Sekimoto}, {and}
  \bibinfo{person}{Satish~V Ukkusuri}.} \bibinfo{year}{2021}\natexlab{}.
\newblock \showarticletitle{Early warning of COVID-19 hotspots using human
  mobility and web search query data}.
\newblock \bibinfo{journal}{\emph{Computers, Environment and Urban Systems}}
  (\bibinfo{year}{2021}), \bibinfo{pages}{101747}.
\newblock


\bibitem[\protect\citeauthoryear{Yabe, Tsubouchi, Shimizu, Sekimoto, and
  Ukkusuri}{Yabe et~al\mbox{.}}{2019}]%
        {yabe2019predicting}
\bibfield{author}{\bibinfo{person}{Takahiro Yabe}, \bibinfo{person}{Kota
  Tsubouchi}, \bibinfo{person}{Toru Shimizu}, \bibinfo{person}{Yoshihide
  Sekimoto}, {and} \bibinfo{person}{Satish~V Ukkusuri}.}
  \bibinfo{year}{2019}\natexlab{}.
\newblock \showarticletitle{Predicting Evacuation Decisions using
  Representations of Individuals' Pre-Disaster Web Search Behavior}. In
  \bibinfo{booktitle}{\emph{SIGKDD}}. \bibinfo{pages}{2707--2717}.
\newblock


\bibitem[\protect\citeauthoryear{Yan, Liao, Cui, Zhang, Hu, and Zhao}{Yan
  et~al\mbox{.}}{2021}]%
        {yan2021multilingual}
\bibfield{author}{\bibinfo{person}{Rui Yan}, \bibinfo{person}{Weiheng Liao},
  \bibinfo{person}{Jianwei Cui}, \bibinfo{person}{Hailei Zhang},
  \bibinfo{person}{Yichuan Hu}, {and} \bibinfo{person}{Dongyan Zhao}.}
  \bibinfo{year}{2021}\natexlab{}.
\newblock \showarticletitle{Multilingual COVID-QA: Learning towards Global
  Information Sharing via Web Question Answering in Multiple Languages}. In
  \bibinfo{booktitle}{\emph{Proceedings of the Web Conference 2021}}.
  \bibinfo{pages}{2590--2600}.
\newblock


\bibitem[\protect\citeauthoryear{Yao, Tang, Wei, Zheng, and Li}{Yao
  et~al\mbox{.}}{2019}]%
        {yao2019revisiting}
\bibfield{author}{\bibinfo{person}{Huaxiu Yao}, \bibinfo{person}{Xianfeng
  Tang}, \bibinfo{person}{Hua Wei}, \bibinfo{person}{Guanjie Zheng}, {and}
  \bibinfo{person}{Zhenhui Li}.} \bibinfo{year}{2019}\natexlab{}.
\newblock \showarticletitle{Revisiting spatial-temporal similarity: A deep
  learning framework for traffic prediction}. In
  \bibinfo{booktitle}{\emph{Proceedings of the AAAI conference on artificial
  intelligence}}, Vol.~\bibinfo{volume}{33}. \bibinfo{pages}{5668--5675}.
\newblock


\bibitem[\protect\citeauthoryear{Yao, Wu, Ke, Tang, Jia, Lu, Gong, Ye, and
  Li}{Yao et~al\mbox{.}}{2018}]%
        {yao2018deep}
\bibfield{author}{\bibinfo{person}{Huaxiu Yao}, \bibinfo{person}{Fei Wu},
  \bibinfo{person}{Jintao Ke}, \bibinfo{person}{Xianfeng Tang},
  \bibinfo{person}{Yitian Jia}, \bibinfo{person}{Siyu Lu},
  \bibinfo{person}{Pinghua Gong}, \bibinfo{person}{Jieping Ye}, {and}
  \bibinfo{person}{Zhenhui Li}.} \bibinfo{year}{2018}\natexlab{}.
\newblock \showarticletitle{Deep multi-view spatial-temporal network for taxi
  demand prediction}. In \bibinfo{booktitle}{\emph{Proceedings of the AAAI
  Conference on Artificial Intelligence}}, Vol.~\bibinfo{volume}{32}.
\newblock


\bibitem[\protect\citeauthoryear{Ye, Hou, Fan, Zhang, Qian, Sun, Peng, Ju,
  Song, and Loparo}{Ye et~al\mbox{.}}{2020}]%
        {ye2020alpha}
\bibfield{author}{\bibinfo{person}{Yanfang Ye}, \bibinfo{person}{Shifu Hou},
  \bibinfo{person}{Yujie Fan}, \bibinfo{person}{Yiming Zhang},
  \bibinfo{person}{Yiyue Qian}, \bibinfo{person}{Shiyu Sun},
  \bibinfo{person}{Qian Peng}, \bibinfo{person}{Mingxuan Ju},
  \bibinfo{person}{Wei Song}, {and} \bibinfo{person}{Kenneth Loparo}.}
  \bibinfo{year}{2020}\natexlab{}.
\newblock \showarticletitle{$alpha$-Satellite: An AI-Driven System and
  Benchmark Datasets for Dynamic COVID-19 Risk Assessment in the United
  States}.
\newblock \bibinfo{journal}{\emph{IEEE Journal of Biomedical and Health
  Informatics}} \bibinfo{volume}{24}, \bibinfo{number}{10}
  (\bibinfo{year}{2020}), \bibinfo{pages}{2755--2764}.
\newblock


\bibitem[\protect\citeauthoryear{Yom-Tov, Lampos, Inns, Cox, and
  Edelstein}{Yom-Tov et~al\mbox{.}}{2022}]%
        {yom2022providing}
\bibfield{author}{\bibinfo{person}{Elad Yom-Tov}, \bibinfo{person}{Vasileios
  Lampos}, \bibinfo{person}{Thomas Inns}, \bibinfo{person}{Ingemar~J Cox},
  {and} \bibinfo{person}{Michael Edelstein}.} \bibinfo{year}{2022}\natexlab{}.
\newblock \showarticletitle{Providing early indication of regional anomalies in
  COVID-19 case counts in England using search engine queries}.
\newblock \bibinfo{journal}{\emph{Scientific reports}} \bibinfo{volume}{12},
  \bibinfo{number}{1} (\bibinfo{year}{2022}), \bibinfo{pages}{1--10}.
\newblock


\bibitem[\protect\citeauthoryear{Zhang, Huang, Xu, Xia, Dai, Bo, Zhang, and
  Zheng}{Zhang et~al\mbox{.}}{2021}]%
        {zhang2021traffic}
\bibfield{author}{\bibinfo{person}{Xiyue Zhang}, \bibinfo{person}{Chao Huang},
  \bibinfo{person}{Yong Xu}, \bibinfo{person}{Lianghao Xia},
  \bibinfo{person}{Peng Dai}, \bibinfo{person}{Liefeng Bo},
  \bibinfo{person}{Junbo Zhang}, {and} \bibinfo{person}{Yu Zheng}.}
  \bibinfo{year}{2021}\natexlab{}.
\newblock \showarticletitle{Traffic Flow Forecasting with Spatial-Temporal
  Graph Diffusion Network}. In \bibinfo{booktitle}{\emph{Proceedings of the
  AAAI Conference on Artificial Intelligence}}, Vol.~\bibinfo{volume}{35}.
  \bibinfo{pages}{15008--15015}.
\newblock


\bibitem[\protect\citeauthoryear{Zhou, Chen, and Lei}{Zhou
  et~al\mbox{.}}{2020a}]%
        {zhou2020not}
\bibfield{author}{\bibinfo{person}{Feng Zhou}, \bibinfo{person}{Tao Chen},
  {and} \bibinfo{person}{Baiying Lei}.} \bibinfo{year}{2020}\natexlab{a}.
\newblock \showarticletitle{Do not forget interaction: Predicting fatality of
  COVID-19 patients using logistic regression}.
\newblock \bibinfo{journal}{\emph{arXiv preprint arXiv:2006.16942}}
  (\bibinfo{year}{2020}).
\newblock


\bibitem[\protect\citeauthoryear{Zhou, Cui, Hu, Zhang, Yang, Liu, Wang, Li, and
  Sun}{Zhou et~al\mbox{.}}{2020b}]%
        {zhou2020graph}
\bibfield{author}{\bibinfo{person}{Jie Zhou}, \bibinfo{person}{Ganqu Cui},
  \bibinfo{person}{Shengding Hu}, \bibinfo{person}{Zhengyan Zhang},
  \bibinfo{person}{Cheng Yang}, \bibinfo{person}{Zhiyuan Liu},
  \bibinfo{person}{Lifeng Wang}, \bibinfo{person}{Changcheng Li}, {and}
  \bibinfo{person}{Maosong Sun}.} \bibinfo{year}{2020}\natexlab{b}.
\newblock \showarticletitle{Graph neural networks: A review of methods and
  applications}.
\newblock \bibinfo{journal}{\emph{AI Open}}  \bibinfo{volume}{1}
  (\bibinfo{year}{2020}), \bibinfo{pages}{57--81}.
\newblock


\end{thebibliography}

\appendix
\newpage
\section{Supplement: Reproducibility}
\subsection{Dataset}
\begin{table}[H]
\vspace{-0.2cm}
\begin{tabular}{cccc}
\toprule[1.5pt]
\addlinespace[1.0pt]
\multicolumn{2}{c}{Location data} & \multicolumn{2}{c}{Web search data} \\ \addlinespace[-2pt]
\cmidrule(r){1-2} \cmidrule(l){3-4} 
\addlinespace[-2pt]
ID1          & thisisfakeid        & ID1       & thisisfakeid            \\
Latitude          & 35.683              & Time      & 2020-03-17 08:32:14     \\
Longitude          & 139.763             & Query     & "Is loss taste covid symptom?"                     \\
Unix time         & 1584390800          &           &                         \\
Date         & 20200317            &           &                         \\ \bottomrule[1.5pt]
\addlinespace[1.0pt]
\end{tabular}
\caption{\label{table_11} Samples of mobility data and web search data.}
\end{table}
\begin{table}[H]
\vspace{-0.7cm}
\begin{tabular}{cccc}
\toprule[1.5pt]
\addlinespace[1.0pt]
\textbf{} & Symptom & \textbf{} & Symptom \\
\addlinespace[-2pt]
\cmidrule(r){1-2} \cmidrule(l){3-4} 
\addlinespace[-2pt]
1              & Abdominal pain             & 23             & Hot flash                        \\ 
2              & Ageusia                    & 24             & Hyperhidrosis                     \\ 
3              & Anosmia                    & 25             & Insomnia                          \\ 
4              & Anxiety                    & 26             & Lethargic                         \\ 
5              & Arthralgia                 & 27             & Loss of appetite                  \\ 
6              & Body ache                  & 28             & Myalgia             \\ 
7              & Chest pain                 & 29             & Nasal dryness                   \\ 
8              & Chest tightness            & 30             & Nausea                     \\ 
9              & Chills                     & 31             & Oropharyngeal pain                           \\ 
10             & Confusion                  & 32             & Pain                \\ 
11             & Cough                      & 33             & Palpitation                         \\ 
12             & Dehydration                & 34             & Pyrexia                      \\ 
13             & Diarrhea                   & 35             & Rash                       \\
14             & Disorientation             & 36             & Rhinorrhea                           \\
15             & Dizziness                  & 37             & Sinusitis                      \\ 
16             & Dyspnea                    & 38             & Sleep disturbance                      \\ 
17             & Ear infection              & 39             & Sneezing              \\ 
18             & Ear pain                   & 40             & Stress                          \\
19             & Eye infection              & 41             & Sweating                           \\ 
20             & Eye pain                   & 42             & Upper Respiratory Tract Infection  \\ 
21             & Fatigue                    & 43             & Vomiting \\ 
22             & Headache                   & 44             & Wheezing                        \\ 
\bottomrule[1.5pt]
\addlinespace[1.0pt]
\end{tabular}
\caption{\label{table_33}Specified COVID-19 related symptoms.}
\vspace{-0.6cm}
\end{table}
This study uses four datasets: $[1]$ mobility data; $[2]$ web search data; $[3]$ infection data; $[4]$ symptom data. 
\begin{itemize}
    \item We aggregate mobility data and  web search data for individuals from the mobile phone data owned by Yahoo Japan Corporation\footnote{ https://about.yahoo.co.jp/en/info/company/}, and feed aggregated mobility (i.e., $\mathbf{E}_{t}$) and web search data (i.e., $\mathbf{H}_{t}$) into our machine learning model. $[1]$ $\mathbf{E}_{t}$ and $[2]$ $\mathbf{H}_{t}$ are accessible at Yahoo! JAPAN R\&D\footnote{https://randd.yahoo.co.jp/en/softwaredata}. \textbf{Please go to YJ Covid-19 Prediction Data}.
    \item $[3]$ infection data and $[4]$ symptom data originate from public resources and can be found under the $data$-$collection$ directory in our Github repository\footnote{ https://github.com/JiaweiXue/MultiwaveCovidPrediction\label{web}}. 
    \item Machine learning codes and implementation results are under the $SAB$-$GNN$ directory in our Github repository.
\end{itemize}

Using these data and codes, readers can fully reproduce the disease prediction results for Tokyo. Besides, readers may use our codes to conduct the disease infection prediction on other cities if their mobility and web search data are available.

\subsection{Data Preprocessing}
\begin{figure}[H]
    \vspace{-0.2cm}
    \centering
    \includegraphics[width=0.7\linewidth]{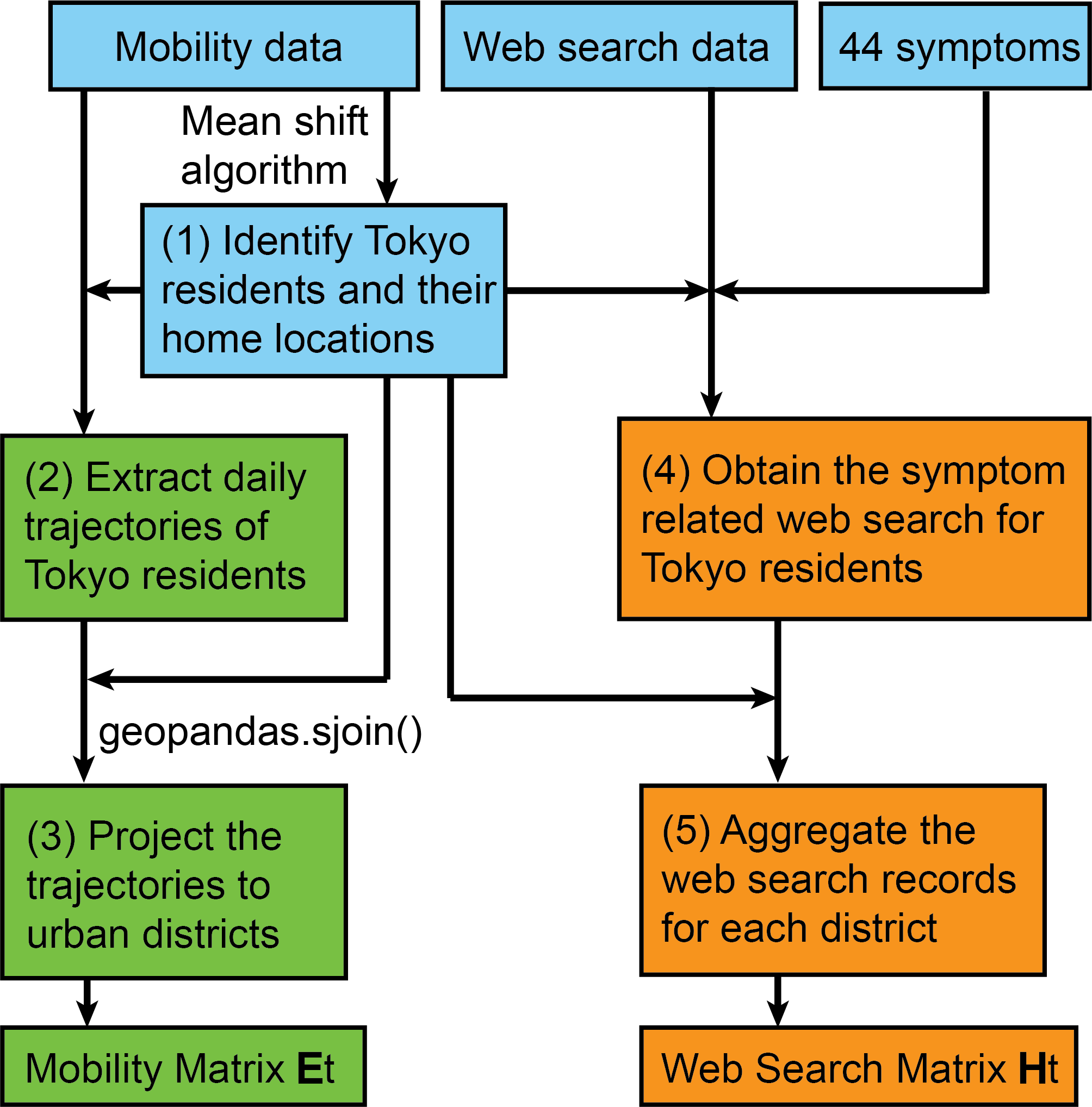}
    \caption{Data preprocessing framework to obtain inputs for machine-learning models.}
    \label{figure11}
    \vspace{-0.25cm}
\end{figure}
Recall that our prediction task requires the standard mobility matrix $\mathbf{E}_{t}$ as well as the web search matrix $\mathbf{H}_{t}$ on the day $t$. We design a data preprocessing framework to convert the raw mobility and web search data to $\mathbf{E}_{t}$ and $\mathbf{H}_{t}$ (Figure~\ref{figure11}). We notice from the raw mobility data that a proportion of user IDs only appear a few times, which indicates that they might be contemporary travelers in Tokyo and may not be strongly related to the inter-city disease spreading in several months, and thus exclude their mobility data in this study. We finally identify 551,745 permanent users in the 23 special wards of Tokyo, which take around 6\% of the total population in these wards. Our mobility and web search data preprocessing procedure is shown as Figure~\ref{figure11}: 
\begin{itemize}
    \item \textbf{Identify Tokyo residents.} We apply the Mean Shift Algorithm \cite{cheng1995mean} to each permanent users' location points during the night hours (from 6:00 PM to 9:00 AM) for 2 weeks starting from Jan. 6, 2020 to estimate the longitude and latitude of their homes using Java (step (1)).
    \item \textbf{Generate mobility feature.} We extract the daily trajectories of these permanent residents using Java (step (2)) and then project the location points along the trajectories to the urban districts using the $spatial-join$ function in $\mathsf{geopandas}$ package in Python and obtain the mobility matrix $\mathbf{E}_{t}$ (step (3)). Note that we mine the daily location trajectory of each user ID and identify cross-district trips with the duration of at least 10 minutes (these trips are referred to as valid trips). 
    \item \textbf{Generate web search feature.} Based on the specified 44 COVID-19 symptoms, we count the number of symptom searches for each permanent resident using Java (step (4)), aggregate them by urban districts (step (5)), and arrive at the web search matrix $\mathbf{H}_{t}$.
\end{itemize}

\begin{algorithm}[h]
\begin{algorithmic}[1]
    \Require
    $\{G_{t}, T-D_{1}+1 \leq t \leq T; I_{t}, T+1 \leq t \leq T+D_{2}\}_{T\in T_{train}}$, where $G_{t} = (V,\mathbf{E}_{t},\mathbf{H}_{t},\mathbf{I}_{t})$.
    \Ensure the SAB-GNN model.
    \For {each epoch} 
        \For {each batch}
           \State $\mathcal{L}_{batch} \leftarrow 0$, $m=0$;
           \For {$T$ in this batch} 
             \State $m$ $\leftarrow m+1$;
             \State \textbf{/* Spatial module */}
             \State Evaluate $\mathbf{X}_{t}^{(k+1)}$ using Eq. (\ref{equ1}), $T-D_{1}+1 \leq t \leq T$;
             \State Generate representations 
             \State $\mathbf{H}_{t}^{S}=\mathbf{X}_{t}^{(K)}$, $T-D_{1}+1 \leq t \leq T$;
             \State \textbf{/* Social awareness recovery module */}
             \State  Update $\mathbf{H}_{t}^{S}$ with $\mathbf{M}_{t,t_{0}}$;
             \State Compute $\mathbf{H}_{t}^{A}$ using Eq. (\ref{equ6});
             \State \textbf{/* Temporal module */}
             \State Feed each row $i$ of $\mathbf{H}_{t}^{A}$ into an LSTM model and 
             \State obtain the hidden states for $t=T+1,T+2,...,T+D_{2}$;
             \State Use a one-layer perceptron to transform 
             \State these hidden states to $D_{2}$ predicted cases $
             \hat{I}_{t,i}$;
             \State Compute the loss $\mathcal{L}_{T}$  using Eq. (\ref{equ7});
             \State $\mathcal{L}_{batch} \leftarrow \frac{m-1}{m}\mathcal{L}_{batch}  + \frac{1}{m}\mathcal{L}_{T}$;
           \EndFor
           \State Use Adam optimizer to update the model parameter.
        \EndFor
    \EndFor
    \end{algorithmic}
    \caption{: Training algorithm of SAB-GNN}
    \label{algo:sab_gnn}
\end{algorithm}

\begin{table*}[t]
\begin{tabular}{ccccc}
\toprule[1.5pt]
\addlinespace[1.0pt]
Study               & Period                & Used features               & Model                 & Publication                 \\ \hline
Chimmula et al. \cite{chimmula2020time}     & Jan. 2020 - Mar. 2020 & Infection                  & LSTM                  & Chaos, Solitons \& Fractals \\
Xiao et al. \cite{xiao2021c}        & Jan. 2020 - Mar. 2020 & Baidu mobility, etc.       & TL                    & AAAI 2021                   \\
Schwabe et al. \cite{schwabe2021predicting}      & Feb. 2020 - Apr. 2020 & Swisscom mobility, etc.    & Hawkes model          & KDD 2021                    \\
Panagopoulos et al. \cite{panagopoulos2021transfer} & Feb. 2020 - May 2020  & Facebook mobility          & MPNN + TL             & AAAI 2021                   \\
Kapoor et al. \cite{kapoor2020examining}       & Mar. 2020 - May 2020  & Google mobility            & GCN + Skip-connection & ArXiv                       \\
Kargas et al. \cite{kargas2021stelar} & Mar. 2020 - Jun. 2020 & Medical data               & Epidemiological model & AAAI 2021                   \\
Gao et al. \cite{gao2021stan}         & Mar. 2020 - Jun. 2020 & Medical data               & GAT + GRU             & JAMIA                       \\
Wang et al. \cite{wang2020examining}       & Mar. 2020 - Aug. 2020 & Medical data               & Ensemble              & IEEE Big Data 2020          \\
Adiga et al. \cite{adiga2021all}       & Aug. 2020 - Jan. 2021 & Infection                  & Ensemble              & KDD 2021                    \\
Chang et al. \cite{chang2021supporting}        & Mar. 2020 - Feb. 2021 & SafeGraph mobility, etc.   & Epidemiological model & KDD 2021\\
Sesti et al. \cite{sesti2021integrating} & Jan. 2020 - May 2021  & Infection                  & GraphLSTM             & ICML 2021 Workshop                       \\ \hline
This study          & Apr. 2020 - May 2021  & Yahoo mobility, web search & Social Awareness GNN  & KDD 2022                 \\ 
\bottomrule[1pt]
\addlinespace[1.0pt]
\end{tabular}
\caption{\label{table_compare}Comparison between existing studies and this study.}
\vspace{-0.5cm}
\end{table*}
\subsection{Training Algorithm}
We present the training algorithm as Algorithm \ref{algo:sab_gnn}. We calculate the gradient and update model parameters for each batch of data. 
\subsection{Hyperparamter Selection}
For SAB-GNN-wsa and SAB-GNN, we try following hyperparameters to identify optimal hyperparameters which are underlined.
\begin{itemize}
    \item Learning rate $\in$ \{1e-1, 1e-2, 1e-3, \underline{1e-4}, 1e-5\}.
    \item Batch size $\in$ \{2, 4, \underline{8}, 16\}.
    \item Dropout rate $\in$ \{0.3, \underline{0.5}, 0.7\}.
    \item $L_{1} \in \{\underline{1},2,3\}$. $L_{1}$: the number of layers in GNN. 
    \item $L_{2} \in \{1,\underline{2},3\}$. $L_{2}$: the number of layers in LSTM.
    \item $n_w$ $\in$ \{1, 3, \underline{8}, 12, 44\}. $n_{w}$: the number of used symptom features  (recall that $\mathbf{h}_{t}^{v_{i}} \in \mathbb{R}^{n_{w}}$).
\end{itemize} 

\subsection{Comparison with Existing Studies}
We present existing COVID-19 infection studies in Table \ref{table_compare} based on prediction period, used feature data, and model architecture. We find that mobility data owned by different companies such as Baidu, Facebook, Google, and Swisscom served as the primary sources of COVID-19 infection prediction before Jun. 2020 when the first COVID-19 outbreaks occurred in many countries. As the pandemic propagated and evolved worldwide with the total infection cases exceeding 530 million as of June 7, 2022 \cite{2022who}, the infection dynamic involved more factors such as virus mutation, mask policy, vaccination, and travel policy changes. 

Our study makes the first attempt to introduce web search data, which is a representative signal of unconfirmed positive cases within the population, into the multiwave infection prediction. Besides, we consider the temporal decay of social awareness of COVID-19 symptoms in our proposed social awareness GNN, to fully leverage the power of web search data. The reason is that people's cognition changed during the two-year pandemic. 

\textbf{Our goal is not to beat all other models.} COVID-19 spreading is a highly uncertain process affected by varying transmission discrepancy, virus mutation, human behavior, and vaccination penetrations in different periods and locations worldwide, which makes a universally best model impossible. \textbf{
Centers for Disease Control and Prevention announced that even the ensemble model was unable to provide reliable predictions all the time\footnote{ https://www.cdc.gov/coronavirus/2019-ncov/science/forecasting/forecasts-cases.html}}. Instead, the main takeaway is that we can leverage web search data to complement existing mobility-based infection prediction to confront the challenges of the multiwave infection uncertainty. 

\subsection{Application on Social Media Data Mining} 
One contribution of this study is the social awareness recovery module (Equations \ref{equ3}, \ref{equ4}, \ref{equ5}). Social awareness depicts the general phenomenon that human perception and attention on events / products / news becomes weaker as time proceeds. Beyond disease prediction, it is promising to apply the social awareness recovery module to social media data from other sources (e.g., Facebook, Twitter, Weibo) in other applications such as predicting the visits to a new location, and the shopping demand of a new product.
\end{document}